\documentclass[reprint,nofootinbib, amsmath,amssymb,prd]{revtex4-1}

\usepackage{hyperref}
\usepackage{graphicx}% Include figure files
\usepackage{dcolumn}% Align table columns on the decimal point
\usepackage{bm}% bold math
\usepackage{xcolor}
\usepackage{xspace}
\usepackage{amsmath}
\usepackage{amssymb}
\usepackage{nicefrac}
\usepackage{multirow}
\usepackage{subcaption}  % preferred

\begin{document}

\title{Spectral distortions in the decaying QCD dark matter scenario} 

\author{Jorge Mastache}
\email{jh.mastache@mctp.mx}
\affiliation{Mesoamerican Centre for Theoretical Physics, Universidad Aut\'{o}noma de Chiapas,  Carretera Zapata Km. 4, Real del Bosque, 29040, Tuxtla Guti\'{e}rrez, Chiapas, M\'{e}xico,}
\affiliation{Secretar\'ia de Ciencia, Humanidades, Tecnolog\'ia e Innovaci\'on, Av. Insurgentes Sur 1582, Colonia Cr\'edito
Constructor, Del. Benito Ju\'arez, 03940, Ciudad de M\'exico, M\'exico} 

\author{Raúl Henriquez-Ortiz}
\email{raul.henriquez@ues.edu.sv}
\affiliation{Facultad de Ciencias en F\'isica y Matem\'aticas, Universidad Aut\'onoma de Chiapas, \\ Carretera Emiliano Zapata Km.\ 8, Rancho San Francisco, Ciudad Universitaria, Ter\'an, Tuxtla Guti\'errez, Chiapas, C.P. 29050, M\'exico,}
\affiliation{Escuela de F\'isica, Facultad de Ciencias Naturales y Matem\'atica, Universidad de El Salvador,\\ final de Av.\ M\'artires y H\'eroes del 30 julio, San Salvador, CP 1101, El Salvador.}

\begin{abstract}
We study the QCD--DM scenario by analyzing the imprint of energy injection from decaying dark-sector particles on the spectral distortions (SDs) of the Cosmic Microwave Background. We adopt a unified framework capable of describing both relativistic and non-relativistic particles, as well as slow and fast decay regimes. Within this approach, we study exponential, power-law, oscillatory, and two-step decay processes, computing the resulting $\mu$-- and $y$--type distortions across the parameter space spanned by the confinement scale $a_c$, decay rate $\Gamma_\chi$, energy-transfer efficiency $\Sigma_\chi$, and velocity dispersion $v_{\chi c}$. We find that power-law, oscillatory, and cascade decays can be effectively mapped onto exponential models after appropriate rescaling. The dominant factors controlling SDs are the decay epoch and lifetime, with $v_{\chi c}$ only becoming relevant in the ultra-relativistic limit. FIRAS observations impose constraints on early energy injection, with $\mu$-type distortions providing the tightest bounds on $\Sigma_\chi$. Simultaneous matching of both $\mu_{\rm firas}$ and $y_{\rm firas}$ breaks the degeneracy between $\Gamma_\chi$ and $\Sigma_\chi$, localizing preferred decay rates around $\Gamma_\chi \lesssim (3.3$-$4.4)\times10^{-3}$~yr$^{-1}$ and $\Sigma_\chi \lesssim 8.5 \times 10^{-4}$ for relativistic particles, and revealing that fast decays with $\Gamma_\chi \gtrsim 6.5$~yr$^{-1}$ or large $\Sigma_\chi$ become observationally negligible. Our extended analysis confirms that steep power-law decays shift distortion epochs earlier and strengthen upper limits on $\Sigma_\chi$. Overall, our results show that CMB spectral distortions are a powerful probe of dark-sector confinement and decay. Future missions, such as \textit{PIXIE} and \textit{PRISM}, could extend current sensitivity by several orders of magnitude and test regions of parameter space that are currently unconstrained, offering a direct observational window into non-standard early-Universe dynamics.
\end{abstract}

\maketitle

%%%% ------------------------------------- %%%
\section{Introduction}\label{introduction} %%%
%%%% ------------------------------------- %%%

% Introduction to DM
In the last couple of decades, the amount and precision of cosmological data have enhanced our understanding of the composition of the Universe. Our canonical cosmological model, the $\Lambda$CDM model, includes two unknown fluids to the standard model of particles: the dark energy, a fluid composing 68\% of the total energy density of the Universe driving its accelerated expansion, i.e., a cosmological constant, $\Lambda$ whose evolution of its equation of state has been heavily constraint just recently \cite{DESI:2025zgx, Planck:2018vyg, eBOSS:2020yzd}; and dark matter, a non-relativistic fluid making the 28\% of the Universe, massive enough to cluster baryonic matter at the large scales, e.g., cold dark matter (CDM) \cite{Bertone:2004pz, Feng:2010gw, Bertone:2018krk}. The dark energy $\Lambda$, the CDM, along with the baryons (filling the remaining 4\% of the Universe), are part of the $\Lambda$CDM model, which can explain the physics of the Universe at different scales and regimes. For example, the Cosmic Microwave Background (CMB) radiation \cite{Riess:1998cb, Perlmutter:1998np, Aghanim:2018eyx, DESI:2025zgx, Planck:2018vyg}, galaxy and quasar surveys \cite{Alam:2020sor, Ahumada:2019vht, Aghamousa:2016zmz, 2009arXiv0912.0201, laureijs2011euclid, BOSS:2012dmf, eBOSS:2020yzd, DESI:2025zgx}, lensing probes \cite{de2013kilo, Heymans:2020gsg} and supernovae catalogs \cite{SDSS:2014iwm}. 

Despite its success, the $\Lambda$CDM model has fundamental caveats explaining the nature of dark matter and dark energy. Moreover, some of the issues in the $\Lambda$CDM model include the too-big-too-fail problem, which refers to the discrepancy between the predicted number of massive satellite galaxies in simulations and the fewer, less massive satellites observed around the Milky Way \cite{Lopez-Corredoira:2017rqn, Ellis:2004rn, Navarro:1996gj, Boylan-Kolchin:2011qkt}; the core/cusp problem: a prediction of cuspy density profiles for the DM distribution in galaxies for which the observation seems to prefer less dense and less cuspy profiles \cite{Albada:1986roa, deBlok:2001hbg, Oh:2010ea, Mastache:2011cn}. Other significant challenges are the $\sigma_8$ and Hubble tensions: the former is the discrepancy between the value of $\sigma_8$ inferred from CMB observations and that obtained from large-scale structure measurements \cite{Lesgourgues:2015wza, Chudaykin:2020acu, DiValentino:2020vvd}, and the later focuses on the discrepancies in the measured expansion rate of the Universe and the apparent evolution of the dark energy equation of state \cite{Aghanim:2018eyx, 
 Alam:2020sor, Ahumada:2019vht, Aghamousa:2016zmz, 2009arXiv0912.0201, laureijs2011euclid, Scolnic:2017caz, Lucca:2020fgp, DESI:2025zgx, Riess:2019cxk, DiValentino:2021izs}.

% Intro to QCD-like dark matter
In the Standard Model (SM) of particle physics, the strong coupling constant of Quantum Chromodynamics (QCD) exhibits asymptotic freedom at high energies \cite{Gross:1973id, Politzer:1973fx}, causing elementary particles such as quarks to behave as weakly interacting, massless, and relativistic entities.  Conversely, at low energies, the increasing strength of the gauge coupling leads to confinement at the QCD scale, resulting in the formation of massive hadrons \cite{Wilczek:1999be, Shifman:1978bx}. Theories that replicate such behaviors, for instance, the SM extended with an extra non-Abelian gauge group, possess a non-trivial structure that may give rise to a dark sector \cite{Kribs:2016cew, Kribs:2010ii, Cline:2013zca, Boddy:2014yra, delaMacorra:2009yb, delaMacorra:2001ay, Appelquist:2015yfa, Kribs:2009fy}. The DM candidates exhibiting confinement and asymptotic freedom are called QCD--DM. Extensions to the SM that incorporate particles in the dark sector with QCD characteristics have been proposed in recent years, for instance, models in the context of twin particles \cite{Chacko:2005pe, Craig:2015pha, Barbieri:2015lqa, Blennow:2010qp, Bittar:2023kdl}, composite or symmetric dark matter scenarios \cite{Antipin:2015xia, Hietanen:2013fya, Detmold:2014qqa, Cline:2021itd}, string theory approaches that naturally generate hidden confining sectors \cite{Svrcek:2006yi, Goodsell:2009xc}; and hidden valley phenomenology \cite{Strassler:2006im, Arkani-Hamed:2008kxc, Han:2007ae}.

% top-down approach/confinement scale
The dynamics of the dark sector depend on the symmetry of the gauge group (e.g., the number of colors), the number of matter fields (e.g., the flavors), and their interactions. Many models can produce particles in the dark sector, and many exhibit physics similar to QCD dynamics; this is, in the QCD--DM scenario, dark fundamental particles have asymptotic freedom at high energies and confinement at low energies. Therefore, we study the cosmological framework from the top-down approach for QCD--DM particles, with no prior expectations on the free parameters the theory may depend on. Instead, we focus on effective features to study the phenomenology of the QCD--DM nature in the cosmological scenario. The natural scenario to study the confinement scale, $\Lambda_c$, in the dark sector is a consequence of the expansion of the Universe because, at the Big Bang, all particles were created at high energy, probably at the grand unification energy scale. As the Universe expands, the energy of the particles decreases until they reach the threshold energy $\Lambda_c$ at a scale factor $a_c$, leading to the formation of gauge-neutral composite particles. 

The study of dark sectors with QCD-like dynamics has garnered significant interest. For instance, investigations into dark matter candidates arising from QCD-DM theories have been conducted, highlighting the potential role of dark pions as thermal relics \cite{Garcia-Cely:2024ivo}. Attempts to constrain $a_c$ include the cosmic microwave background radiation (CMB) \citep{Mastache:2019bxu}, Big Bang Nucleosynthesis \cite{Giovanetti:2021izc, Mastache:2013iha}, and large-scale structure \cite{Bringmann:2018jpr, Mastache:2019bxu}.

% DM particles properties: mass
The dark-sector hadronization will have a collection of dark bound states that, depending on the complexity of the undergoing model, are likely to have dark particles with different mass, lifetime, electric charge, etc. The mass of the bounds states is bigger than the sum of its former particles, proportional to $\Lambda_c$, and depends on the strength of the dark gauge force; examples of this is that at QCD confinement scale, $\Lambda_{\rm QCD} \simeq 332$ MeV \cite{ParticleDataGroup:2018ovx}, gauge invariant states are created forming color-neutral particles, i.e., mesons and baryons, with non-perturbative masses, e.g., $m_\pi \simeq 134.98$ MeV, and $m_{n}\simeq 939.57$MeV, which are much larger than the sum of the masses of their constituted quarks ($m_u\simeq 2.3$MeV,  $m_d\simeq  4.8$MeV) and proportional to $\Lambda_{\rm QCD}$.  Beyond QCD, the gauge-mediated SUSY breaking model shows similar non-perturbative mass generation mechanisms \cite{Giudice:1998bp}. The spectrum and decay properties of such dark hadrons can impact cosmology and be probed via displaced vertices or resonance searches at colliders \cite{Cohen:2020afv, Cline:2021itd, Bai:2013xga}. Recent studies using parton-shower simulations have shown that dark hadronization can give rise to emerging particle jets depending on the lifetimes and mass gaps of the dark bound states \cite{Schwaller:2015gea, Cacciapaglia:2020vyf, Bellazzini:2015cgj}. These analyses reinforce the role of non-perturbative dynamics in shaping the phenomenology of QCD-DM.
 
% DM particles properties: velocity
The velocity dispersion of the particles at the confinement scale, $v_{ic}$, is a measurement of the binding force and relevant at the cosmological scale because it could imply a not negligible free--streaming scale \cite{Mastache:2019bxu, Bringmann:2005pp} preventing the small structure from forming at the early Universe that can lessen the missing satellite problem \cite{Bullock:2017xww, Lovell:2011rd}. This phenomenological approach has also been studied inside high densities structures (e.g., galaxies halos) where the energy density increases with decreasing radius in dark matter halos, and the confinement scale is constrained using the rotation curves of galaxies of low surface brightness \cite{delaMacorra:2009yb, delaMacorra:2011df, Mastache:2011cn, deBlok:2001hbg}. These analyses have been particularly useful for testing the viability of QCD-DM models with suppressed small-scale structure. Therefore, QCD--DM candidates could help lessen the caveats of the $\Lambda$CDM scenario by introducing a physical mechanism for suppressing substructure formation and modifying halo profiles on galactic scales.

% Stable, unstable DM; % QCD example
The average lifetime is another important aspect of particle physics; it provides insights into the fundamental dynamic of the dark sector because it is proportional to the coupling constant and the particle mass. On the one hand, some dark particles may be electric neutral and stable particles that could be good candidates for cold DM (CDM) or warm DM, depending on the velocity dispersion $v_{\rm dmc}$ and the confinement epoch $a_c$ it could describe one candidate or another \cite{Mastache:2019gui, Mastache:2019bxu, Chu:2011be, Boehm:2013jpa}. For instance, the value of $a_c$ had been constrained to $a_c \leq 2.66 \times 10^{-6}$ for CDM ($v_{\rm dmc} = 0$) using Planck, SNIa, and BAO data \cite{Mastache:2019bxu}. On the other hand, an unstable (short-lived) dark particle can also be expected to decay after the confinement scale $a_c$. In QCD, mesons are particles that have short life span compared to baryons: e.g., a proton has a long lifetime, since no decay has been detected, its lifetime is of $\mathcal{O}(10^{32})$ years, neutron decays into protons with a lifetime of $\mathcal{O}(10^3)$ s, and charged pions has a lifetime of $\sim 2.6 \times 10^{-8}$ s. The average lifetime of particles can range from fractions of a second to billions of years; therefore, with no prior knowledge of the dark gauge group and since the average lifetime of DM particles has not been measured, no constraints on its value can be made.
Nevertheless, indirect methods have been used to constrain the lifetimes of unstable DM candidates. For example, analyses of the extragalactic gamma-ray background suggest that dark matter must have lifetimes longer than $10^{26}$ seconds to avoid excess gamma-ray flux \cite{Cohen:2016uyg, Slatyer:2016qyl, Essig:2013goa}. Similarly, CMB measurements strongly constrain decaying dark matter due to its impact on ionization history and the anisotropy power spectrum \cite{Poulin:2016anj, Diamanti:2013bia}.

% mean lifetime \Gamma_\chi
We assume that two kinds of particles were formed after the confinement scale was reached. The first is stable, with a long life, and has a mass $m_{\rm dm}$, is electrically neutral, and resembles the fiducial CDM. The second would have mass $m_\chi$, a relatively short life ($\Gamma_\chi$) compared to the CDM particle that eventually will decay. Although many dark particles can compose one kind of this family of particles, each can be described as a single fluid, allowing for the study of its cosmological signatures for simplicity. Previous studies constrain the lifespan of a DM based only on the assumption that it decays into relativistic particles, finding that $\Gamma_\chi \leq 0.01 \, {\rm Gyr^{-1}}$ (95\% CL) \cite{Ichiki:2004vi, DeLopeAmigo:2009dc, Audren:2014bca}. Further analyses have put lower limits on the dark matter lifetime, generally in the range of $\tau \sim (1-5) \times 10^{28}$ seconds, depending on the decay channels and dark matter mass \cite{Cohen:2016uyg, Slatyer:2016qyl, Poulin:2016anj, Ando:2015qda}. These bounds apply to scenarios in which DM is stable on cosmological timescales and decays after recombination. In contrast, the decaying component in the QCD-DM framework is not the present-day DM, but a transient species in the early Universe that injects energy just before confinement.

% photon creation
It is interesting to consider the production of monochromatic photons in the final state after the DM decay because the search for photon production, such as gamma rays, has been extensively pursued as an indirect probe of DM \cite{Fermi-LAT:2015att, Topchiev:2017xfp, Galper:2014pua}. For the case of direct production of a pair of monochromatic photons, the mass $m_\chi$ can be connected to the energy of the photon final states by $E_\gamma = f_{\rm eff} m_\chi/2$, when $f_{\rm eff} = 1$ is the maximum amount of energy that the photon can have. The constant $f_{\rm eff}$ accounts for any other decaying channel or if the photon is a secondary product of intermediate decaying states. For instance, axion is a well-motivated CDM candidate which can weakly couple to photons \cite{Raffelt:2006cw, Mirizzi:2006zy, Matsuki:1990mf, CAST:2004gzq, ADMX:2019uok, Sikivie:1983ip}. Experiments such as the Axion Dark Matter Experiment have been designed to detect such interactions \cite{ADMX:2018gho}.

% Creating SD
The photon produced in the QCD--DM decay interacts with the surrounding photons and matter, causing energy transfer and contributing to the CMB photons, depending on the cosmological parameters of the model. In the early Universe, photons and baryons are tightly coupled, behaving as a single viscous fluid close to thermal equilibrium due to Compton, Bremsstrahlung, and double Compton scattering processes, which isotropize the photon-baryon fluid, creating a blackbody spectral distribution. However, early energy injection, as the decay scenario of the QCD--DM model, can disrupt thermal equilibrium, causing small departures from the blackbody distribution; these deviations are the spectral distortions (SD) of the CMB \cite{Chluba:2011hw}. Such distortions provide an independent and complementary probe to study the nature of dark matter. For instance, energy injection from decaying dark matter can lead to observable $\mu$-type and $y$-type distortions in the CMB spectrum \cite{Acharya:2019uba}. Additionally, the dissipation of acoustic waves in the early Universe can also generate spectral distortions, offering insights into primordial density fluctuations \cite{Chluba:2012gq}. We aim to find constraints on the confinement phenomenological parameters ($a_c$, $v_{\chi c}$, $\Gamma_\chi$, and $\Sigma_\chi = f_{\rm eff} f_\chi$) using the SD.

% Introduction to $\mu$ and $y$ distortions
At early times ($3 \times 10^{5} \lesssim z \lesssim 2 \times 10^{6}$), interactions such as double Compton scattering and Bremsstrahlung become inefficient at maintaining thermal equilibrium, leading to a departure from a perfect blackbody spectrum energy computed with a chemical potential, the resulting departure is known as the $\mu$-type SD \cite{Sunyaev:1970eu, Burigana:1991eub}. At later times ($z \lesssim 1 \times 10^{4}$), when Compton scattering becomes inefficient at redistributing energy, energy injections produce $y$-type SD, analogous to the thermal Sunyaev-Zeldovich effect observed in clusters of galaxies \cite{Zeldovich:1969ff}.

Several astrophysical and cosmological processes have been studied through their contributions to SD. For instance, reionization and structure formation introduces $y$-type distortions due to the heating of electrons during these epochs \cite{Hu:1994bz}. Cosmological inflation can lead to $\mu$-type distortions through the dissipation of acoustic waves at small scales \cite{Chluba:2012we, Mastache:2023cha, Henriquez-Ortiz:2022ulz}. Decaying or annihilating relic particles inject energy into the CMB, potentially causing both $\mu$ and $y$-type distortions, depending on the epoch of energy release \cite{Chluba:2011hw, Chluba:2013pya, Poulin:2016anj, Kogut:2011xw, Chluba:2013vsa, Khatri:2012tv, Bolliet:2020ofj, Acharya:2019uba}. Additionally, the cosmological recombination process itself introduces subtle spectral features known as the cosmological recombination radiation \cite{Rubino-Martin:2006hng}. Processes such as the adiabatic cooling of electrons and baryons also contribute to spectral distortions \cite{Hu:1994bz, Khatri:2012tv, Chluba:2011hw}.

Observations from the COBE/FIRAS instrument have constrained these distortions, with limits of $|\mu| \leq 9 \times 10^{-5}$ and $|y| \leq 1.5 \times 10^{-5}$ at 95\% CL \cite{Fixsen:1996nj}. Since many theoretical models predict smaller spectral distortions, more precise experimental measurements are necessary to detect these signals. Proposed missions like the Primordial Inflation Explorer (PIXIE) \cite{Kogut:2011xw}, its enhanced version Super-PIXIE \cite{Chluba:2019nxa}, the Polarized Radiation Imaging and Spectroscopy Mission (PRISM) \cite{PRISM:2013ybg}, and the ESA's Voyage 2050 program \cite{Chluba:2019kpb} aim to explore spectral distortions with expected sensitivities of $\sigma(\mu) \simeq 7.7\times10^{-9}$ and $\sigma(y) \simeq 1.6\times10^{-9}$ for Super-PIXIE. Even alternative configurations have been proposed with $\mathcal{O}(1000)$ times better sensitivity \cite{Fu:2020wkq}. Such sensitivity would be enough to ruled out several DM models.

This paper is organized as follows: in Sec.~\ref{sec_2}, a review of the parameters that help to effectively describe the QCD--DM scenario; in Sec.~\ref{sec_3}, the details of the theory of CMB spectral distortions from heat injection is presented from the QCD--DM scenario We present the main results and details in Sec.~\ref{sec_4}. Finally, the conclusions are shown in Sec.~\ref{sec.conclusion}.

%%%%%%%%%%%%%%%%%%%%%%%%%%%%%%%%%%%%%%%%%%%%%%%%%%%%%%%%%%%%%%%%%%
\section{Fluid approximation for QCD--Dark Matter} \label{sec_2}%%
%%%%%%%%%%%%%%%%%%%%%%%%%%%%%%%%%%%%%%%%%%%%%%%%%%%%%%%%%%%%%%%%%%
% QCD-DM definition
QCD dark matter (QCD--DM) refers to a class of dark matter candidates whose gauge coupling constant becomes weaker at high energies, exhibiting a behavior analogous to the asymptotic freedom and confinement properties of QCD. By hypothesis, this dark gauge group is not part of the SM of particle physics but arises from an additional non-Abelian gauge sector. These sectors can exhibit rich dynamics similar to QCD and are theoretically motivated by a range of frameworks. Some frameworks include: SM extensions with new confining gauge groups that introduce additional gauge sectors \cite{Kribs:2016cew, Kribs:2010ii, Cline:2013zca, Boddy:2014yra, delaMacorra:2009yb, delaMacorra:2001ay, Appelquist:2015yfa, Kribs:2009fy, Agrawal:2014aoa, Baek:2013qwa, Blennow:2010qp}; composite and symmetric DM models consisting of stable bound states of new strongly interacting particles \cite{Antipin:2015xia, Hietanen:2013fya, Detmold:2014qqa, Cline:2021itd}; mirror or twin Higgs models, which address the hierarchy problem via a mirror sector \cite{Chacko:2005pe, Craig:2015pha, Barbieri:2015lqa, Blennow:2010qp, Bittar:2023kdl};  Hidden valley–like frameworks, characterized by a hidden sector with a confining gauge group and weak couplings to the SM, typically mediated by heavy connector particles \cite{Strassler:2006im, Arkani-Hamed:2008kxc, Han:2007ae};  and string-theory motivated hidden sectors, in which compactification schemes or intersecting branes give rise to dark confining sectors with minimal couplings to the visible sector \cite{Svrcek:2006yi, Goodsell:2009xc}. 

% top-down approach / confinement scale
These frameworks highlight the broad theoretical appeal of QCD--DM, offering diverse dynamics and multiple pathways for DM to emerge from strongly coupled dynamics. The cosmological phenomenology of such models can depend on key parameters, including the confinement scale ($\Lambda_\chi$), the number of dark hadrons, the nature of interactions between the dark sector and the SM fields, and the mass scale of the resulting dark hadrons. While the space of possible particle physics realizations in the dark sector is extensive, many models lead to qualitatively similar outcomes—namely, the formation of composite dark hadrons with varying masses and lifetimes. 
    
Given the extension of models, we adopt a phenomenological approach to describe QCD--DM within the cosmological framework, as developed in \cite{Mastache:2019bxu}. In this approach, the dark sector is initially composed of massless, relativistic particles, for instance, dark quarks ($\chi_q$) at energies above a critical confinement scale $\Lambda_c$, corresponding to a cosmological scale factor $a_c$. For $a < a_c$, the dark quarks behave as radiation. 

% DM particles properties: mass
As the Universe expands, the energy of particles redshifts inversely with the scale factor. When the temperature of the dark sector falls below the confinement scale $\Lambda_c(a_c)$ marking the onset of hadronization. Below this threshold ($a < a_c$), the dark sector binds its elementary constituents dark quarks ($\chi_q$) into a spectrum of dark bound states ($\chi_h$) with a range of masses, lifetimes, and potentially electric charges. Based on dimensional analysis in spontaneous symmetry breaking and confinement theories, the masses of these dark bound states are set by the strength of the underlying gauge interaction and scale with the confinement energy. At leading order, a bound state composed of $n$ dark quarks has a mass $m_\chi \propto n \chi_q \Lambda_c$, with the binding energy dominating the total mass. In the absence of specific constraints on the dark sector gauge theory, it is also possible for the masses of the bound states to significantly exceed the confinement scale, $m_\chi \gg \Lambda_c$. Such scenarios give rise to exotic phenomena known as quirks \cite{Albouy:2022cin}. However, this case fall beyond the scope of the phenomenological model considered in this work.

% DM particles properties: velocity
Since the masses of the bound states are determined by the strength of the dark gauge interaction, each species of dark hadron (denoted as $i$-particle) acquires an initial velocity dispersion, $v_i(a_c) = v_{ic}$, which serves as an indirect probe of the underlying binding force. The kinetic properties of these particles can be characterized by their momentum, $p_i = \gamma_{ic} m_i v_{ic}$, where $\gamma_{ic}$ is the Lorentz factor. A state is relativistic if its momentum significantly exceeds its rest mass (i.e., $p_i \gg m_i$), and non-relativistic if $p_i \ll m_i$. This initial velocity dispersion is a key parameter in determining the free-streaming scale and the subsequent impact of the dark sector on structure formation \cite{Mastache:2019bxu}.

% Stable, unstable DM; mean lifetime \Gamma_\chi
The minimum number of fundamental dark fields required to form a bound state is two, consisting of a dark quark and its corresponding opposite dark-flavor charge, $\chi_q$ and $\chi_{\overline{q}}$. The resulting $\chi_q \chi_{\overline{q}}$ bound states are expected to organize into multiplets under a dark-flavor SU($N$) symmetry, analogous to QCD mesons. While multi-$\chi_q$ singlet states may also exist (e.g., baryon-like configurations), the larger number of $\chi_q$ particles implied a lower probability of creation. Following the confinement phase transition, we assume two primary dark bound states are generated. The first class consists of stable, electrically and dark-flavor neutral, non-relativistic particles behaving as dark matter that rapidly become non-relativistic after formation (e.g., for $v_{dmc} = 0$ such as cold dark matter, CDM). The second class, denoted as $\chi$, corresponds to unstable, color-neutral bound states and an average lifetime characterized by the decay constant $\Gamma_\chi$. These particles are assumed to decay into Standard Model particles, either directly emitting photons or producing them through subsequent decay chains. Although each class may comprise a spectrum of meson- or baryon-like states, we model them as effective single-fluid components for simplicity. This approximation assumes that one dominant decay mode characterizes the behavior of the unstable species and can be described by a representative $\Gamma_\chi$.

% a_c constrains
We adopt the constraints on the confinement scale factor $a_c$ derived in \cite{Mastache:2019bxu, delaMacorra:2018zbk, Torres:2025qko}, where the QCD-DM model was tested against observational data from Planck 2018, Type Ia Supernovae (SNIa), and Baryon Acoustic Oscillations (BAO). Based on this, we define three benchmark scenarios based on the velocity dispersion of the stable DM component at the transition: (i) for particles born non-relativistic ($v_{\rm dmc} = 0$), resembling cold dark matter (CDM), the upper bounds are $a_c \leq 2.66 \times 10^{-6}$ at $1\sigma$; (ii) for particles formed with relativistic velocities ($v_{\rm dmc} = 1/\sqrt{2}$) the bounds tighten to $a_c \leq 4.59 \times 10^{-7}$ at $1\sigma$ CL. (3) It was shown that a transition at $a_c = 3.18 \times 10^{-7}$ with $v_{\rm dmc} = 0$ can effectively mimic the free-streaming behavior of a thermal 3 KeV WDM particle, reproducing the same cutoff scale in the matter power spectrum \cite{Mastache:2019bxu}. Such a scenario is consistent with Lyman-$\alpha$ forest data and suppresses the halo mass function at scales $\mathcal{O}(10^7)\, M_\odot$, offering a potential resolution to the missing satellite problem. These constraints are particularly relevant because the values are close to the redshift of the $\mu$-distortion era, $z_\mu \simeq 2 \times 10^6$ (corresponding to a scale factor $a_\mu \simeq 5 \times 10^{-7}$). In this regime, spectral distortions in the CMB can emerge. Energy injections occurring before $a \lesssim a_{\rm th} \simeq 5 \times 10^{-7}$  are efficiently thermalized with photons, erasing any spectral distortions. 
    
% QCD example
To illustrate the feature discussed above regarding the mass and mean lifetime of dark sector particles, we take QCD as an example. Both electrically charged mesons and electrically neutral baryons are formed during the process of QCD confinement. These hadrons are color-neutral yet exhibit a broad range of masses and lifetimes. As an example of a decaying particle, we take the neutral pion ($\pi^0$), which has a mass of $m_{\pi^0} = 134.98$ MeV and is composed of a quark-antiquark pair (up and anti-up or down and anti-down), with $m_u = 2.16$ MeV and $m_d = 4.67$ MeV. Notice that its mass is not the sum of its constituent quark masses, but rather of the order of the QCD confinement scale $\Lambda_{\rm QCD}$, with binding energy contributing significantly. The pion mean lifetime is extremely short, $\tau_{\pi^0} = 8.43 \times 10^{-17}$ s~\cite{ParticleDataGroup:2022pth}. As an example of a neutral, stable particle, we take the neutron, which formed from one up quark and two down quarks and has a mass of $m_n = 939.57$ MeV, much greater than the sum $m_u + 2m_d$. Its mean lifetime is $\tau_n = 878$ s~\cite{ParticleDataGroup:2022pth}. While protons are stable on cosmological timescales, their lifetime is $\tau_p > 3.6 \times 10^{29}$ years~\cite{ParticleDataGroup:2022pth}. The confining gauge theory can generate particles spanning many orders of magnitude in a lifetime while their masses remain proportional to the underlying confinement scale. This behavior serves as a useful analogy for understanding similar dynamics in QCD-DM models.

%%%%%%%%%%%%%%%%% ----- FIG photon creation  ----- %%%%%%%%%%%%%%%%%%%%
\begin{figure}[t] 
\begin{center}
    \includegraphics[width=0.235\textwidth]{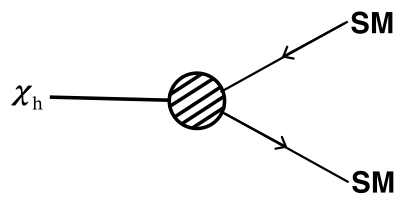}
    \caption{Schematic diagrams representing the decay of a dark hadron $\chi_h$ into two Standard Model (SM) particles through an effective interaction vertex. The hatched circle denotes the effective operator or mediator responsible for the decay process.}
    \label{fig:schematic}
\end{center}
\end{figure}
%%%%%%%%%%%%%%%%% ----- FIG photon creation  ----- %%%%%%%%%%%%%%%%%%%%

Experimental searches for DM, including QCD-DM, focus on detecting weak interaction signals with SM particles, given that DM is known to interact only very weakly with ordinary matter. One possible indirect signature is the production of photons following the decay of unstable DM particles \cite{Blennow:2010qp, Baek:2013qwa, Agrawal:2014aoa, Cline:2013zca, Strassler:2006im, Arkani-Hamed:2008kxc, Han:2007ae, Svrcek:2006yi, Goodsell:2009xc, Cline:2021itd, Craig:2015pha, Chacko:2005pe, Barbieri:2015lqa, Bittar:2023kdl}. A schematic representation of this interaction is shown in Figure~\ref{fig:schematic}. In scenarios where the decay produces two monochromatic photons, their energy is given by $E_\gamma = f_{\rm eff} m_\chi / 2$, where $m_\chi$ is the mass of the decaying particle and $f_{\rm eff} < 1$ accounts for energy loss into other decay products or intermediate states. Since the precise particle physics model for QCD--DM is not specified in this study, the phenomenological parameter $f_{\rm eff}$ incorporates the effects of different decay channels transferring energy to photons.

% Creating SD
A fraction of the energy from the decaying bound states in the dark sector manifests as extra radiation in the early Universe that can interact with the photon-baryon plasma via Compton or Thomson scattering, transferring energy to the CMB photons. Such interactions can perturb the blackbody spectrum, resulting in SD.

%%%%%%%%%% ------ DENSITY EVOLUTION FIGURE -----  %%%%%
\begin{figure}[t]
\begin{center}
    \includegraphics[width=0.49\textwidth]{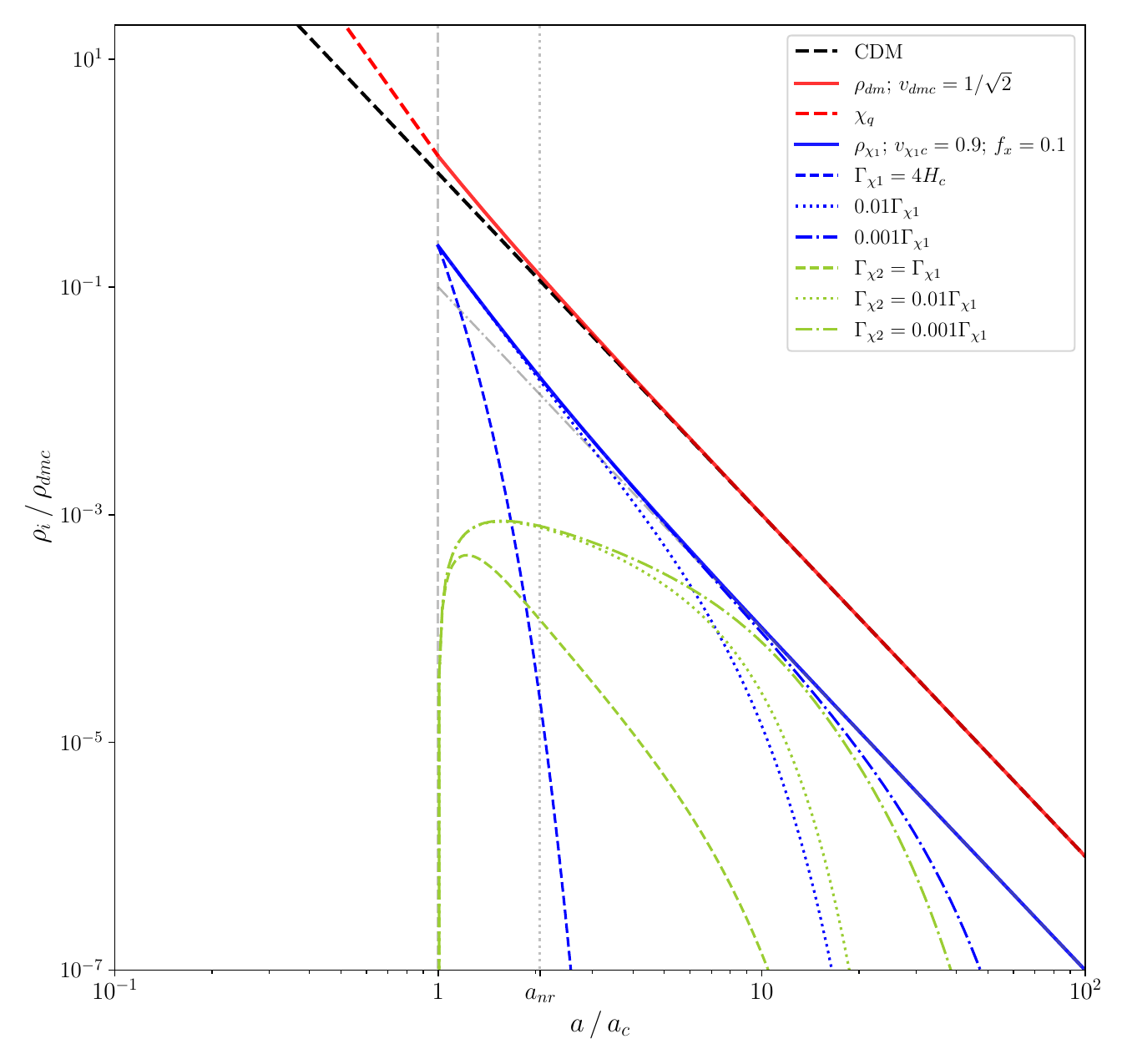}
        \caption{Evolution of the energy density of dark-sector components in the QCD--DM scenario, normalized to $\rho_{dmc}$ and $a_c$. The red lines show the stable QCD--DM component before and after confinement, while the blue lines depict the decaying species $\chi_1 (v_{\chi 1}=0.9, f_\chi=0.1)$ for different decay rates $\Gamma_{\chi_1}=[1.0,\,0.01,\,0.001]\,4H_c$. Green lines correspond to the daughter particles $\chi_2$ for fixed $\Gamma_{\chi_1}=0.01\,\Gamma_{\chi 1}$ and varying $\Gamma_{\chi_2}$. The dashed black line represents standard CDM, and vertical markers indicate the confinement and non-relativistic transition epochs.}
    \label{fig:rho_evolution_qcd_dm}
\end{center}
\end{figure}
%%%%%%%%%% ------ DENSITY EVOLUTION FIGURE -----  %%%%%

%%% ----------------------------- %%%
\subsection{Equation of evolution} %%
%%% ----------------------------- %%%
We assume that the former particles, $\chi_q$, are relativistic and massless. In contrast, after the transition ($a > a_{c}$), the $i$-dark bound states are formed with a non-perturbative mass and a velocity dispersion $v_{ic}$. The stable DM particle is assumed to be cold at $a_c$, $v_{dm}(a_c) = v_{dmc} = 0$, unless specified otherwise. No constraints on $v_{\chi c}$ or $\Gamma_\chi$ can be made on the decaying state; depending on the value of $v_{\chi c}$, the $\chi$ particle may still be in a relativistic or non-relativistic regime. Throughout this work, any quantity with the subindex $c$ is evaluated at the transition epoch $a = a_c$. 

To describe the evolution of the dark matter energy density, we follow the analytical framework given in \cite{Mastache:2019gui, Mastache:2019bxu}. In this approach, the evolution of the velocity of a {\it i}-decoupled massive particle in an expanding universe is obtained exactly as a function of the scale factor. Starting from the relativistic relation
\begin{eqnarray}\label{eq:vel_bdm}
  v_i(a)  &=& \frac{\gamma_{ic} v_{ic}(a_c/a)}{\sqrt{1 + \gamma^2_{ic} v^2_{ic}(a_c/a)^2}} \; ,
\end{eqnarray} 
where we have used the Lorentz factor $ \gamma_{ic} \equiv (1 - v_{ic}^2)^{-1/2}$. The equation of state parameter is defined as $\omega_i = v_i^2/3$. With this definition and the equation of state for the pressure $P = \omega \rho$, the continuity equation $\dot{\rho} = -3H(\rho + P)$ can be integrated analytically, yielding a general expression for energy density, $\rho(a)$ that is valid both in the relativistic and non-relativistic regimes. The energy density for the stable dark matter and decaying $\chi$ particles is
\begin{eqnarray}
  \rho_{\rm dm}(a) &=& \rho_{\rm dmc} \left( \frac{a}{a_{c}} \right)^{-4}\left( \frac{v_{\rm dmc}}{v_{\rm dm}(a)} \right) , \label{eq:dm_energy_density}\\
  \rho_{\rm \chi}(a) &=& \rho_{\rm \chi c} \left( \frac{a}{a_{c}} \right)^{-4}\left( \frac{v_{\chi c}}{v_\chi(a)} \right) e^{-\Gamma_\chi (t - t_c)}  , \label{eq:chi_energy_density}
\end{eqnarray}\label{eq:bdmrho}
This formulation correctly reproduces the expected limits $\rho \propto a^{-4}$ in the relativistic regime and $\rho \propto a^{-3}$ if the particle is non-relativistic. This approximation offers a physically transparent description of the dark matter energy density evolution, eliminating the need to solve the complete Boltzmann hierarchy. 

Starting from the expression for the dark matter energy density at the confinement scale, Eq.~\eqref{eq:bdmrho}, and using the non-relativistic limit $a_c \ll a_0$ to relate the present-day and confinement velocities ($v_{\rm dm0} \approx v_{\rm dmc}\gamma_{\rm dmc}(a_c/a_0)$), we can express the density of the unstable component $\chi$ in terms of the total dark matter density today. By introducing the fractional abundance $f_{\chi} = \rho_{\chi c}/\rho_{\rm dmc}$ and accounting for the redshifting of the energy density, the velocity evolution $v_\chi(a)$, and the decay factor $e^{-\Gamma_\chi (t-t_c)}$, we obtain 
\begin{equation}
    \rho_\chi (a) = f_{\chi} \gamma_{\chi c} \, \rho_{\rm dm0} \left( \frac{a}{a_0}\right)^{-4} \left(\frac{a_c}{a_0}\right) \left(\frac{v_{\chi c}}{v_\chi(a)}\right) e^{-\Gamma_\chi(t-t_c)}\; .
    \label{eq:rho_x}
\end{equation}
We expected $f_\chi$ to be $\mathcal{O}(<10)$, taking the meson-to-baryon fraction in QCD as reference \cite{Braun-Munzinger:2003pwq, Andronic:2017pug}. No constraints can be made for the velocity dispersion of $\chi$, $v_{\chi c}$, in contrast with $v_{dmc}$, which is constrained by large-scale structure formation. 

We define a scale–dependent Lorentz factor as
$$
\gamma_\chi(a)\equiv \frac{1}{\sqrt{1-v_\chi(a)^2}}\,.
$$
Using the kinematic identity $p/m=\gamma v$ and the redshifting of the momentum $p(a)=p_c\,(a_c/a)$, we can prove that
$$
\gamma_{\chi c}\,\frac{v_{\chi c}}{v_\chi(a)} \;=\; \frac{a}{a_c}\,\gamma_\chi(a)\,.
$$
Substituting the last expressión into Eq.\eqref{eq:rho_x} yields the compact form
\begin{equation}
\;\rho_\chi(a) \;=\;
f_\chi\,\rho_{\rm dm0}\,\gamma_\chi(a)\,
\left(\frac{a}{a_0}\right)^{-3}\,
e^{-\Gamma_\chi (t-t_c)}\; .
\label{eq:rho_x_gammaofa}
\end{equation}
This expression makes transparent that the density redshifts as matter,
\(\propto a^{-3}\), modulated by the relativistic factor \(\gamma_\chi(a)\)
and the decay factor \(e^{-\Gamma_\chi (t-t_c)}\). It reproduces the correct limits: $\gamma_\chi(a)\!\to\!1$ (non-relativistic), therefore, $\rho_\chi\propto a^{-3}$; while for ultra-relativistic particles \(\gamma_\chi(a)\propto a^{-1}\),
so \(\rho_\chi\propto a^{-3}a^{-1}=a^{-4}\).

Constraints in the parameter space $a_{c}-v_{c}$ had been established for the bound DM  model (see Fig.4 in \cite{Mastache:2019bxu}), which form part of the scenario that QCD--DM can be described, we aim to find constraints on the $a_{c}-v_{c}$ parameters space using spectral distortions.

%%%%%%%%%% ------ ENERGY INJECTION -----  %%%%%
\begin{figure*}[t]
    \centering
    % First image
    \begin{subfigure}[b]{0.48\textwidth}
        \centering
        \includegraphics[width=\linewidth]{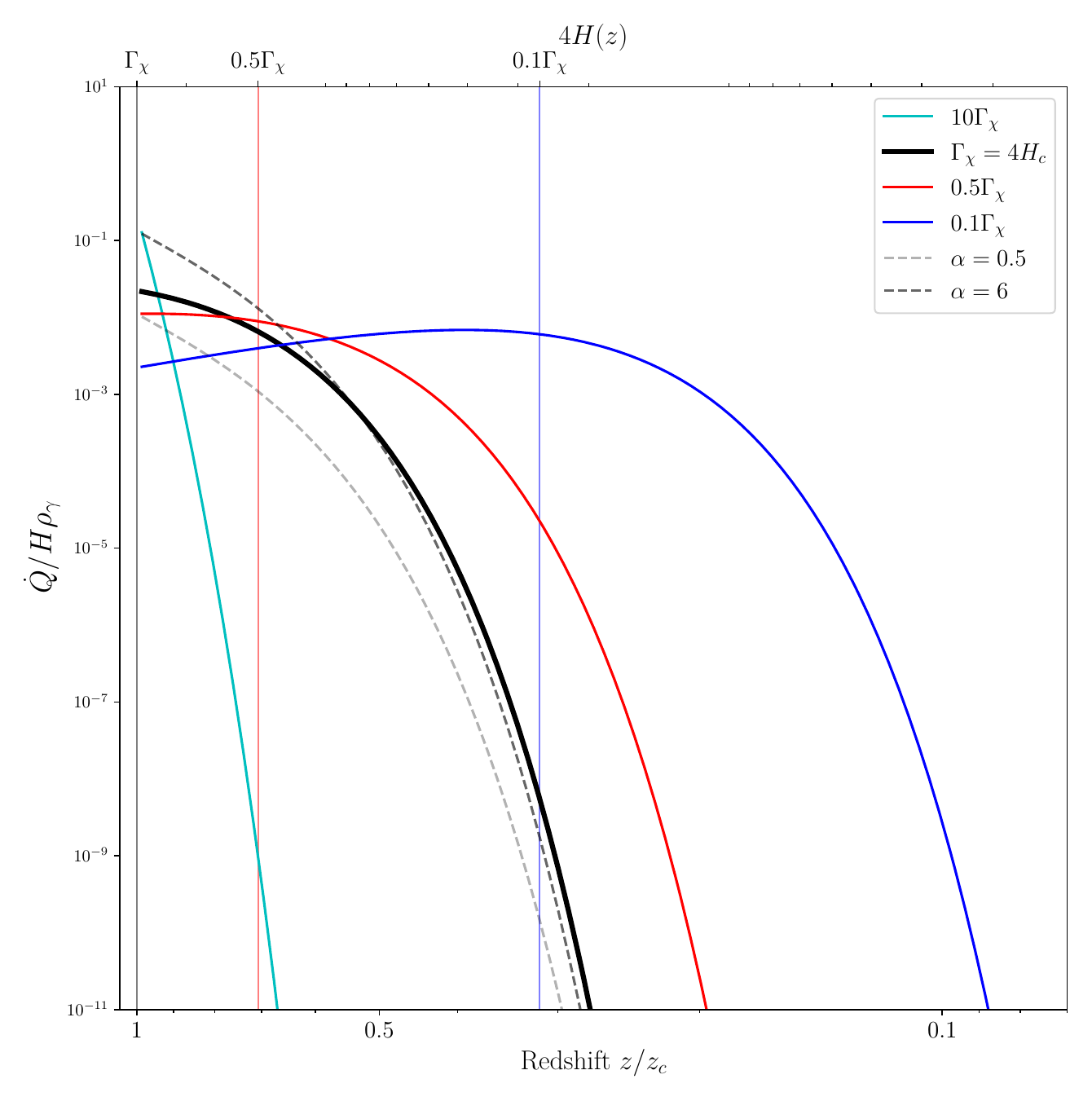}
        \caption{Panel A: Exponential and power-law decay.}
        \label{fig:caseA}
    \end{subfigure}
    \hfill
    % Second image
    \begin{subfigure}[b]{0.48\textwidth}
        \centering
        \includegraphics[width=\linewidth]{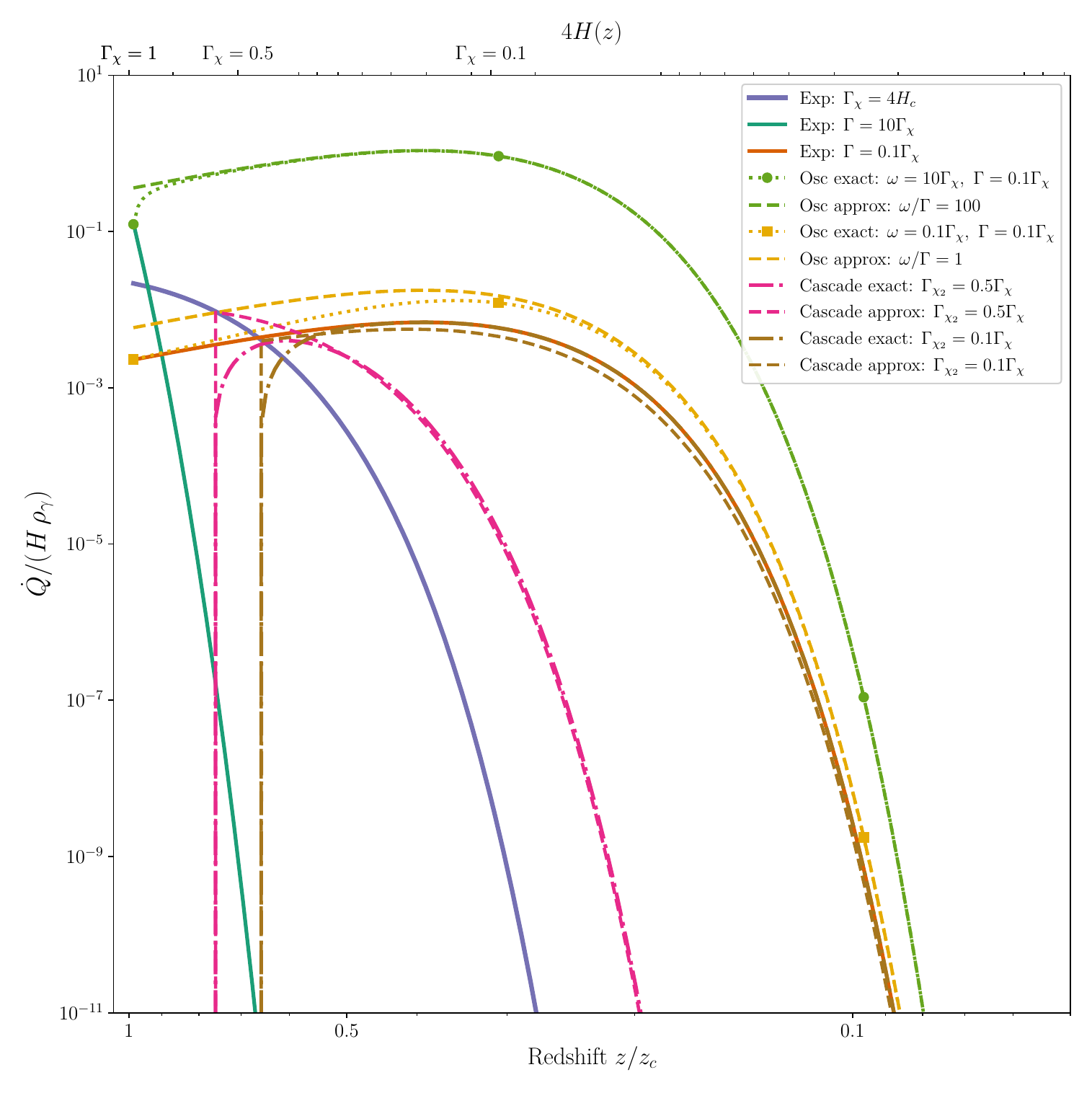}
        \caption{Panel B: Oscillatory and two-step decay.}
        \label{fig:caseB}
    \end{subfigure}
    \caption{Energy-injection rate $\dot Q/(H\rho_\gamma)$ normalized to the Hubble expansion as a function of $z/z_c$, where $z_c$ is stablish at the beginning of the $\mu$-era. \textbf{(a)} Exponential energy-injection histories for decay rates $\Gamma=[10,\,1,\,0.5,\,0.1]\Gamma_{\chi}$ and power-law indices $\alpha=\{0.5,6\}$ with fixed $\Gamma_{\chi} = t_\chi =4H_c$. Vertical dashed lines mark epochs when $\Gamma = 4H$, corresponding to the onset of strong decay. \textbf{(b)} Comparison of different decay mechanisms: exponential (reference), oscillatory, and two-step decay, showing that an exponential decay can effectively approximate the latter. For oscillatory decay, Case 1 (green) corresponds to a fast decay ($\omega/\Gamma=100$), and Case 2 (yellow) to a moderate decay ($\omega/\Gamma=1$). For the cascade decay, Case 3 (magenta) corresponds to $\Gamma_{\chi_2}=0.5\,\Gamma_{\chi_1}$, while Case 4 (brown) to $\Gamma_{\chi_2}=0.1\,\Gamma_{\chi_1}$, and both cases have $\Gamma_{\chi1} = \Gamma_\chi$. See Table~\ref{tab:decay_osc_2step} for the resulting spectral distortions.}
    \label{fig:dQdt_all}
\end{figure*}
%%%%%%%%%% ------ ENERGY INJECTION -----  %%%%%

%%%%%  ----------------------------------------------------------------------     %%%%%%
\subsection{Energy injection from decaying particles}  \label{subsec:energy_deposition}  %%%%
%%%%%  ----------------------------------------------------------------------     %%%%%%
The evolution of the photon energy density is $\dot{\rho}_\gamma + 4H\,\rho_\gamma = q(z)$, where $q \equiv \dot{Q}/V$ represents the energy per unit volume and time deposited into the photon fluid.  Throughout this work we assume a homogeneous energy injection, that is, the spatial average of the source term within $V$ is equal to the local rate of energy injection, $\frac{1}{V}\frac{dQ}{dt} = \frac{\partial^2E_{\rm inj}}{\partial V \partial t}$, and all times are measured with respect to $t_c$.

For $a < a_c$, no energy is transferred to photons, i.e.\ $(d^2E/dVdt)_{\rm inj}=0$. Dark hadronization occurs when the Universe reaches the scale factor $a_c$ (at time $t_c$); at that moment, the number density of the $\chi$ particles starts to decrease. After the decay of the $\chi$ particles, the energy injection per unit volume into the background photons is modulated by a deposition efficiency factor $f_{\rm eff}(z,E) \in [0,1]$, which encodes the fraction of the injected energy that is effectively deposited into heating or ionizing the photon--electron plasma at a given redshift $z$ \cite{Slatyer:2016qyl, Poulin:2016anj, Furlanetto:2009uf}.

A distinctive feature of our framework is that the particles can be either relativistic or non-relativistic after the confinement transition. The natural evolution of their energy density self-consistently accounts for the time-dilation factor of the decaying population, allowing us to explore both short- and long-lived decay regimes in a unified description.

Several functional forms can describe particle decays.  The most common and widely studied is the exponential decay law, but other behaviors may occur depending on the underlying physics. In this work we consider different decay prescriptions: exponential, power-law, oscillatory, and cascade (two-step) decays.

\subsubsection{Exponential decay}
In the exponential case, the comovil number density of particles evolves as $N(t) = N_0\, e^{-\Gamma_\chi t}$. Deriving the continuity equation using this expression we notice that the energy injection into the photon fluid is
\begin{equation}
    \left. \frac{\partial^2 E}{\partial V\,\partial t}\right|_{\rm inj}^{\chi} = f_{\rm eff}\, \Gamma_\chi\, \rho_\chi,
\label{eq:exponential_injected_energy}
\end{equation}
where $N_0(a)$ is the initial number density at $t_c$.

The Hubble rate sets the competition between decay and expansion at the confinement scale, $H_c \equiv H(a_c)$. From Fig.~\ref{fig:dQdt_all}, we notice that the energy injection reaches a maximum when the decaying rate is $\Gamma_\chi \simeq 4H_c$, values plotted by the horizontal lines. Therefore, we define fast decay particles if $\Gamma_\chi > 4H_c$, and slow decay if $\Gamma_\chi < 4H_c$, i.e., interaction faster/slower than the expansion rate, where $4H_c$ is the dilution term given by the expansion in a radiation-dominated Universe in the continuity equation. Eqns.~\eqref{eq:exponential_injected_energy} and \eqref{eq:rho_x} describe both relativistic and non-relativistic $\chi$ particles as well as fast and slow lifetimes within a single fluid framework.

\subsubsection{Power-law decay}
In some contexts, such as the decay of heavy or metastable particles, the comovil number density follows a power-law rather than an exponential behavior, $N(t) = N_0(a)\,(1 + \gamma t)^{-\alpha}$, where $\gamma$ is a characteristic timescale regulating the function near $t=t_c$, and $\alpha>0$ controls the decay rate. 
For $0<\alpha<1$, the total injected energy can diverge at long times and must be handled carefully; $\alpha=1$ corresponds to a maximal-entropy case, and $\alpha>1$ describes faster, well-behaved decays. 
The parameter $\gamma$ is related to the particle lifetime and is approximately the inverse of the decay rate. 
The injected energy density then reads
\begin{equation}
\left. \frac{\partial^2E}{\partial V\,\partial t}\right|_{\rm inj}^{\chi} = 
f_{\rm eff} \frac{\alpha \gamma}{1+\gamma t}\, \rho.
\end{equation}

\subsubsection{Oscillatory decay}

In models with mass mixing or neutrino–like states, the decay probability may oscillate, leading to a combined exponential–oscillatory behavior for the comoving number density, $N(t) = N_0(a)\,e^{-\Gamma_\chi (t-t_c)}\cos[\omega (t-t_c)]$.

The corresponding energy injection rate is
\begin{equation}
    \left.\frac{dE}{dV\,dt}\right|_{\rm inj}^{\chi} = f_{\rm eff}\,\rho(t)\,[\,\Gamma_\chi+\omega\,\arctan(\omega t)\,],
\end{equation}
where the second term represents an additional modulation proportional to the oscillation frequency $\omega$. The ratio between the energy injection of oscillatory over the exponential case help us define de function 
\begin{equation}
    \tilde{f}_{\chi}(t) = f_\chi\left( 1+\frac{\omega}{\Gamma_\chi}\,\arctan(\omega t) \right)
\end{equation}
this acts as a modulation factor that for early times $(\omega t\!\ll\!1)$, $\arctan(\omega t)\!\simeq\!\omega t$ and $\tilde{f_\chi} \to f_\chi$; for late times $(\omega t\!\gg\!1)$, $\arctan(\omega t)\!\to\!\pi/2$, simplifying
\begin{equation}
    \tilde{f}_\chi \simeq f_\chi \Big( 1 + \frac{\pi}{2}\frac{\omega}{\Gamma_\chi} \Big)
\end{equation}
Hence, the oscillatory decay can be mapped onto an exponential decay with a rescaled $\tilde{f}$. When $\omega/\Gamma_\chi\ll1$, both descriptions are indistinguishable; for $\omega/\Gamma_\chi\!\gtrsim\!1$, the mapping above provides an accurate approximation.  

This equivalence is illustrated in Fig.~\ref{fig:caseB}, where oscillatory (dotted) and its exponential approximation (dashed) curves coincide within numerical accuracy. We show two regimes: a fast-oscillating configuration with $\omega/\Gamma_\chi = 100$ (yellow) and a moderated one with $\omega/\Gamma_\chi = 1$ (green). For both regimes, the exponential mapping captures the amplitude and temporal evolution of the heating rate, with minor deviations once the oscillatory modulation saturates. This confirms that a single exponential with a renormalization factor can effectively represent an oscillatory decay.

\subsubsection{Two-step (cascade) decay}
The last case we consider is a two-step decay chain,
\[
\chi_1 \xrightarrow{\Gamma_{\chi_1}} \chi_2 \xrightarrow{\Gamma_{\chi_2}} \gamma,
\]
The particle $\chi_1$, with decay rate $\Gamma_{\chi_1}$, produces an intermediate (secondary) particle $\chi_2$ that subsequently decays into photons with rate $\Gamma_{\chi_2}$. The intermediate particle $\chi_2$ acts as an energy reservoir before decaying into photons. In the comoving rest frame, the number of intermediate particles evolves as
\begin{equation}
N_{\chi_2}(t) = 
\frac{\Gamma_{\chi_1}}{\Gamma_{\chi_2}-\Gamma_{\chi_1}}\, N_{\chi_1c}
\left(e^{-\Gamma_{\chi_1} t} - e^{-\Gamma_{\chi_2} t}\right)
+ N_{\chi_2c}\, e^{-\Gamma_{\chi_2} t},
\label{eq:part2_number_serie}
\end{equation}
where the first term represents $\chi_2$ particles produced by the decay of $\chi_1$, and the second term corresponds to any preexisting population.

The energy density of $\chi_2$ can be decomposed into three contributions
\begin{equation}
\rho_{\chi_2} =  \rho_{\chi_2}^{\rm P} -  \rho_{\chi_2}^{\rm D} +  \rho_{\chi_2}^{\rm ext},
\label{eq:rho_x2}
\end{equation}
representing production ($P$), decay ($D$), and externally created (${\it ext}$) components. 
The total energy injection rate from the chain decay chain is then
\begin{equation}
    \left. \frac{dE}{dV\,dt}\right|_{\rm inj}^{\rm tot} = \left. \frac{dE}{dV\,dt}\right|_{\rm inj}^{\rm P} - \left. \frac{dE}{dV\,dt}\right|_{\rm inj}^{\rm D} + \left. \frac{dE}{dV\,dt}\right|_{\rm inj}^{\rm ext},
    \label{eq:energy_inj_x2}
\end{equation}
Finally, the net energy injection to background photons from the two-step decay is
\begin{equation}
    \left. \frac{dE}{dV\,dt}\right|_{\rm inj}^{\rm tot} = f_{\rm eff} \left( \rho_{\chi_2}\Gamma_{\chi_2} - \rho_{\chi_1}\Gamma_{\chi_1}\frac{E_2}{E_1} \right).
\label{eq:energy_inj_serie}
\end{equation}
where $f_{\rm eff}$ is the effective energy injected into the background photon from the decay series. 

%\paragraph{Exponential mapping.}
If either $\chi_1$ or $\chi_2$ decays rapidly, the behavior can be approximated by a single exponential decay described by Eq.~\eqref{eq:exponential_injected_energy}, after appropriate parameter redefinitions. Depending on the hierarchy between $\Gamma_{\chi_1}$ and $\Gamma_{\chi_2}$, several limiting cases arise:
\begin{enumerate}
    \item $\Gamma_{\chi_1} \gg 1$ and $\Gamma_{\chi_2} \ll 1$: \\
    $N_{\chi_2}(t) \approx N_{\chi_1c}\, e^{-\Gamma_{\chi_2} t}$.
    \item $\Gamma_{\chi_1} \gg 1$ and $\Gamma_{\chi_2} \gg 1$ with $\Gamma_{\chi_2} > \Gamma_{\chi_1}$: \\ 
    $N_{\chi_2}(t) \approx (N_{\chi_1c} \Gamma_{\chi_1}/(\Gamma_{\chi_2}-\Gamma_{\chi_1})) e^{-\Gamma_{\chi_1} t}$.
    \item $\Gamma_{\chi_2} \gg 1$ and $\Gamma_{\chi_1} \ll 1$: \\
    $N_{\chi_2}(t) \approx (N_{\chi_1c}\Gamma_{\chi_1}/\Gamma_{\chi_2}) e^{-\Gamma_{\chi_1} t}$.
\end{enumerate}
For long-lived species ($\Gamma_{\chi_1},\Gamma_{\chi_2}\!\ll\!1$), the energy injection from $\chi_2$ becomes relevant once $\rho_{\chi_2}>0$, neglecting external sources.  
The corresponding delay is approximately
\begin{equation}
    \delta t_{\chi_2} \simeq \frac{1}{\Gamma_{\chi_2}-\Gamma_{\chi_1}} \ln\!\left( \epsilon \frac{\Gamma_{\chi_2}}{\Gamma_{\chi_1}}\right),
\end{equation}
where $\epsilon$ condensate information about the ratio of the energy density between $\chi_1$ and $\chi_2$ given by the particular decaying channel. The effective injection begins at 
$\tilde{a}_c = a_c \left( 1 + \frac{\delta t}{t_c} \right)^{1/2}$, assuming a radiation-dominated epoch.  
A modulation factor is then defined as 
$\tilde{f_\chi} \simeq f_\chi \left(\frac{a_c}{\tilde{a}_c}\right)e^{-\Gamma_{\chi_2}\delta t}$.

The cascade decay can thus be represented by an effective exponential process characterized by $(\tilde{a}_c,\,\tilde{f})$. This approximation reproduces the exact solutions with high precision, as shown in Fig.~\ref{fig:caseB}, where each cascade curve shares its color with its exponential counterpart. 

Hence, cascade and oscillatory decays can be expressed in a unified parameter space \{$\Gamma_\chi,\, \Sigma_\chi,\,a_c,\, v_c$\}, enabling direct comparison across different decay mechanisms.

The Fig.~\ref{fig:caseB} shows the comparison for the two-step (cascade) decay in pink and brown dot-dashed lines, the exponential approximation (dashed with same colors) obtained by shifting the onset scale factor ($\tilde{a}$) and applying a modulation factor $\tilde{f}_\chi$ closely matches the exact energy injection Eq.\eqref{eq:energy_inj_serie}. This mapping effectively reproduces the delayed energy release expected in cascade processes while preserving the same overall redshift dependence as the single exponential model.

%%%%% -------------- TABLE APPROXIMATIONS ------------%%%%%
\begin{table*}[t]
\centering
\setlength{\tabcolsep}{5pt}
\renewcommand{\arraystretch}{1.2}
\begin{tabular}{lcccccc}
Decay Type & Case & $\mu_{\rm exact}$ & $\mu_{\rm approx}$ & $y_{\rm exact}$ & $y_{\rm approx}$ & Ratio $(\mu, y)$ \\ 
\hline
\multirow{2}{*}{Oscillatory decay} 
& 1 & $1.369$ & $1.392$ & $2.16\times10^{-6}$ & $2.16\times10^{-6}$ & (1.017, 1.001) \\
& 2 & $1.556\times10^{-2}$ & $2.264\times10^{-2}$ & $3.38\times10^{-8}$ & $3.52\times10^{-8}$ & (1.455, 1.041) \\
\hline
\multirow{2}{*}{Cascade decay} 
& 3 & $9.12\times10^{-4}$ & $2.95\times10^{-3}$ & $1.04\times10^{-23}$ & $8.19\times10^{-24}$ & (3.239, 0.791) \\
& 4 & $4.59\times10^{-3}$ & $6.28\times10^{-3}$ & $1.37\times10^{-8}$ & $9.50\times10^{-9}$ & (1.367, 0.695)
\end{tabular}
\caption{Comparison between exact and exponential-approximation results for the oscillatory and cascade decay scenarios. $\mu$ and $y$ denote the spectral distortion amplitudes, and the ratios correspond to $\mathrm{exact}/\mathrm{approx}$. The correspondence Case 1-4 are defined in the caption of Fig.~\ref{fig:dQdt_all}.
}
\label{tab:decay_osc_2step}
\end{table*}
%%%%% -------------- TABLE APPROXIMATIONS ------------%%%%%
%%%%%  -----------------------------------------------------------     %%%%%%
\subsection{Regimes and heating rate evolution}  \label{subsec:regimes}  
%%%%%%%%%  ---------------------------------------------------------     %%%%%%
%\paragraph{Kinematic regimes.}
After formation at $a_c$, the species $\chi$ can emerge relativistic if $v_{\chi c} \to 1$. 
As the Universe expands, the momentum redshifts as $p_\chi\propto a^{-1}$ and the velocity decreases according to Eq.~\eqref{eq:vel_bdm}. The transition to the non–relativistic regime occurs when $p_\chi=m_\chi$, i.e.\ at $v_\chi=1/\sqrt{2}$, which defines $a_{\chi{\rm nr}} \;=\; a_c\, v_{\chi c}\,\gamma_{\chi c}$.

For $a\ll a_{\chi{\rm nr}}$, $\gamma_\chi \simeq v_{\chi c}(a_c/a)$ and $\chi$ behaves effectively as radiation; for $a\gg a_{\chi{\rm nr}}$, $\gamma_\chi \to 1 $ and $\chi$ behaves as non–relativistic matter (see Fig.~\ref{fig:rho_evolution_qcd_dm}). This kinematic evolution is encoded in Eq.~\eqref{eq:rho_x_gammaofa} through the factor $\gamma_\chi(a)$, the effective decay width in the cosmological frame is reduced by time dilation. For instance, in the exponential decay $\Gamma_{\rm eff}(a) = \Gamma_\chi \gamma_\chi^{-1}(a)$.

Hence, while $\chi$ is highly relativistic, $\gamma_\chi^{-1} \ll 1$ and decays are suppressed; 
as the Universe expands, $\gamma_\chi$ decrees and the decay rate turns on; once $\chi$ is non–relativistic ($\gamma_\chi\to 1$), the decay proceeds at the rest–frame rate $\Gamma_\chi$.

%\paragraph{Dynamical regimes.}
The energy–injection history reduces to the commonly used short–lived non-relativistic approximation (cf.\ \cite{Lucca:2019rxf, Poulin:2016anj, Chluba:2013pya}). Using the notation $\Sigma_\chi \equiv f_\chi f_{\rm eff}$ and the confinement redshift $z_c$, one obtains
\begin{equation}
\frac{dQ}{dz} ~\simeq~ 
\frac{\Sigma_\chi}{\gamma_{\chi c}}\,
\frac{\Omega_{dm0}\,\rho_{\rm cr0}\,(1+z)^2\,\Gamma_\chi}{H(z)}\,
\exp\!\left[-\tilde{\Gamma}_\chi \!\left(\frac{z_c^2}{z^2}-1\right)\right],
\label{eq:dQ/dz_shortlife}
\end{equation}
where $\tilde{\Gamma}_\chi \equiv \Gamma_\chi\, t_{\rm eq}\,(a_c/a_{\rm eq})^{2}$ is a convenient dimensionless parametrization for the redshift evolution.\footnote{If a different normalization is used in the rest of the paper, $\tilde{\Gamma}_\chi$ can be redefined accordingly without loss of generality.}

The exponential in Eq.~\eqref{eq:rho_x_gammaofa} cannot be neglected, and the full time–dilated evolution must be retained. Within the fluid approximation, this holds at all scale factors without requiring an explicit transfer function for $\chi$ (in contrast to \cite{Poulin:2016anj}). 
This framework can naturally lead to cases where the energy release extends into (or across) the $\mu$–era, with observational consequences for spectral distortions.

%\paragraph{Parameter ranges used.}
For $(a_c, v_c)$ we follow the $1\sigma$–$2\sigma$ regions from \citep{Mastache:2019bxu}(Planck+SNe~Ia+BAO). We span $v_c$ from cold to warm/hot kinematics.  For $\Sigma_\chi$ we adopt the range informed by \citep{Lucca:2019rxf}, which provides $95\%$ bounds on the decaying–DM fraction vs.\ lifetime at PIXIE sensitivity, corresponding roughly to $\Sigma_\chi\simeq 10^{-7}$–$10^{-2}$. In the next section, we compute SD amplitudes for these benchmarks using the formalism of Sec.~\ref {sec_3}. The upper bound on the lifetime of dark particles \cite{Ichiki:2004vi, DeLopeAmigo:2009dc, Audren:2014bca} is larger than the space parameter for our interest, one in which dark particles decay in an epoch where we can create SD. 

%%%%%%%%%%%%%%%%%%%%%%%%%%%%%%%%%%%%%%%%%%%%%%%%%%%%%%%%%%%%%%%%%%%
\section{Spectral distortions from dark sector}\label{sec_3} %%%%%%
%%%%%%%%%%%%%%%%%%%%%%%%%%%%%%%%%%%%%%%%%%%%%%%%%%%%%%%%%%%%%%%%%%%

%%%% ------------- ac - vxc ---------------%%%%
\begin{figure*}[t!]
    \centering
    % First image
    \begin{subfigure}[b]{0.49\textwidth}
        \centering
        \includegraphics[width=0.99\linewidth]{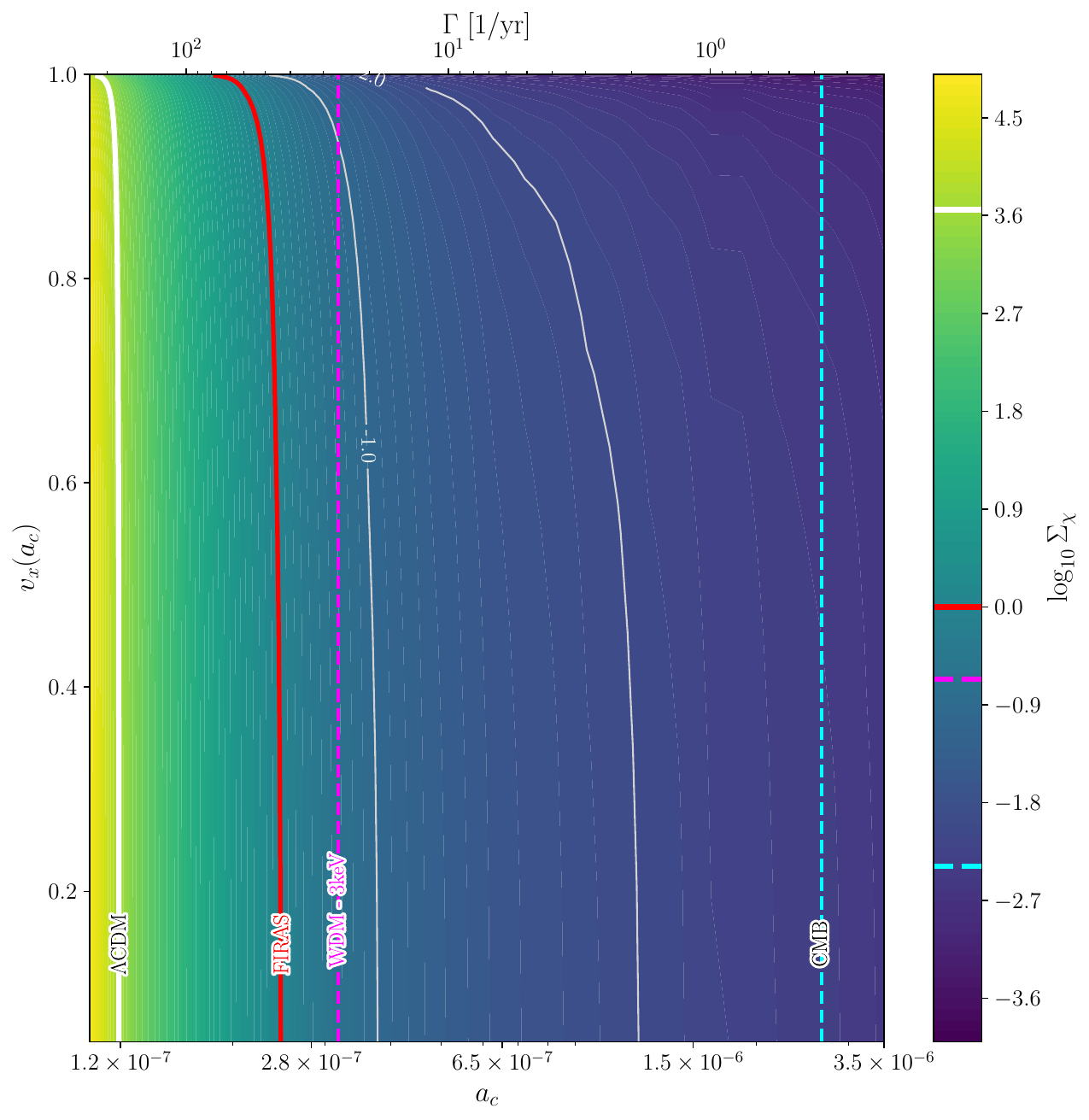}
        \caption{$\mu$-spectral distortions.}
        \label{fig:log10_sigma_mu}
    \end{subfigure}
    \hfill
    % Second image
    \begin{subfigure}[b]{0.49\textwidth}
        \centering
        \includegraphics[width=0.99\linewidth]{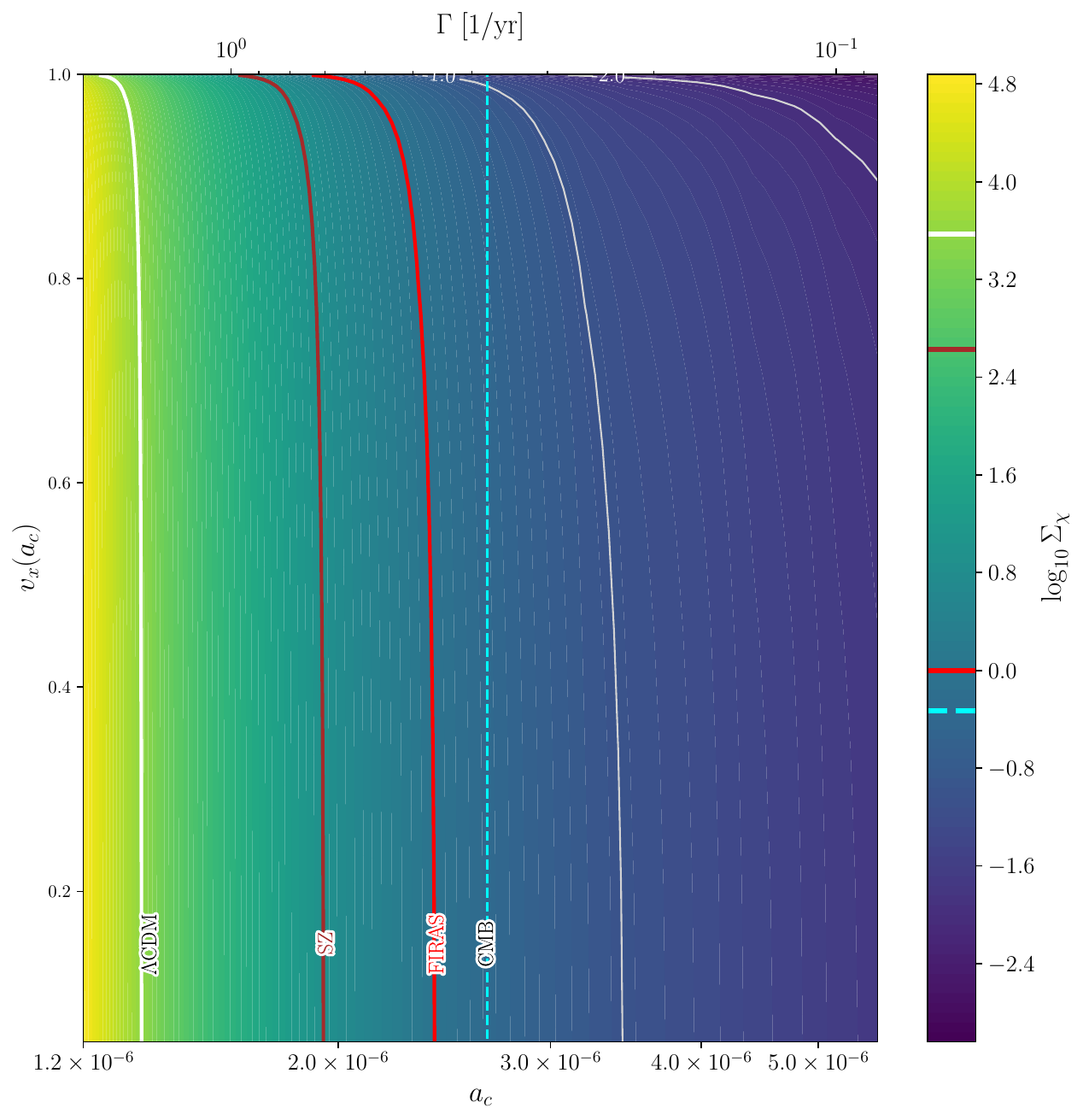}
        \caption{$y$- spectral distortions.}
        \label{fig:log10_sigma_y}
    \end{subfigure}
    \caption{ Contour plot of the constraints on $\Sigma_\chi$ in the $a_c$–$v_{\chi c}$ parameter space. The left and right panels show isolines of constant $\mu=\mu_{\mathrm{FIRAS}}$ and $y=y_{\mathrm{FIRAS}}$, respectively. The cyan vertical line marks the upper bound on $a_c$ from CMB data, while the red isoline corresponds to the FIRAS limits for $\mu_{\rm firas}$ and $y_{\rm firas}$. The white isoline indicates the $\Lambda$CDM prediction ($\mu_{\Lambda\mathrm{cdm}}$, $y_{\Lambda\mathrm{cdm}}$). The magenta vertical line denotes the QCD–DM case with $a_c = 3.18\times10^{-7}$, equivalent to a 3 keV warm–DM model, and the brown isoline shows the late-time $y$-distortion reference value ($y=y_{\mathrm{sz}}$). Positive $\Sigma_\chi$ values represent the inverse scaling factor applied to the $\mu$- or $y$-type distortions relative to the FIRAS reference. }
    \label{fig:sigma_x_constraints}
\end{figure*}
%%%% ------------- ac - vxc ---------------%%%%

The SDs of the CMB could encode information about a wide range of physical processes that perturb the thermal equilibrium of the photon distribution. Energy injection mechanisms, such as the decay, modify the CMB frequency spectrum at specific epochs in cosmic history.  Therefore, departures from a perfect blackbody can serve as a sensitive probe of new physics beyond the standard cosmological model.

In this section, we briefly review the basic formalism to compute $\mu$-- and $y$--type distortions, which we later apply to the QCD--DM scenario. The total distortion of the photon intensity spectrum, $\Delta I$, can be written at first order in terms of the temperature shift $\Delta T$, the Compton $y$--distortion, and the chemical potential $\mu$--distortion \citep{Lucca:2019rxf}:
\begin{equation}\label{eq:sd1}
\Delta I (\nu) ~\approx~ \frac{ \Delta T}{T}\, G (\nu) + y\, Y_{\mathrm{SZ}} (\nu)  + \mu\, M_{\mathrm{SZ}} (\nu)\,.
\end{equation}

Each of these terms can be determined using the Green’s function approach introduced in~\cite{Fu:2020wkq}. The first term in Eq.~\eqref{eq:sd1} describes a temperature shift that changes the blackbody spectrum proportionally to
\begin{equation}
    G(\nu)~=~T \,\frac{\partial B}{\partial T}  \,,
    \qquad
    B(\nu)~=~\frac{2 h \nu^{3}}{c^{2}\left(e^{x}-1\right)}\,,
\end{equation}
where $B(\nu)$ is the Planck function.  The spectral shape functions multiplying $y$ and $\mu$ parametrize the out-of-equilibrium effects of Compton scattering and the chemical potential modification of the photon distribution \cite{Zeldovich:1969ff}, and are defined as
\begin{subequations}
\begin{eqnarray}
    Y_{\mathrm{SZ}}(\nu) &\simeq& G(\nu)\,[\,x \coth(x)-4\,]\,, \\
    M_{\mathrm{SZ}}(\nu) &\simeq& G(\nu)\,\left(0.4561 - \frac{1}{x} \right)\,,
\end{eqnarray}
\end{subequations}
where $x \equiv h\nu / k_{\!B}T$ is the dimensionless frequency.

In what follows, we focus on $y$– and $\mu$–type distortions arising from energy injection due to decaying dark matter. A first-order estimate of the total distortion can be obtained by the comovil integral of the effective energy release \cite{Chluba:2013dna, Chluba:2013vsa, Chluba:2022xsd}
\begin{eqnarray}
    \mu &\simeq& 1.401 \int d \ln (1+z) \mathcal{J}_\mu(z) \frac{Q(z)}{H(z) \rho_\gamma(z)}, \label{eq:sd2} \\
    y &\simeq& \frac{1}{4} \int d \ln (1+z) \mathcal{J}_y(z) \frac{Q(z)}{H(z) \rho_\gamma(z)} \label{eq:sd3}
\end{eqnarray}
Here, $\mathcal{J}_\mu(z)$ and $\mathcal{J}_y(z)$ are thermalization windows that weight the epoch at which the injected energy is converted into a $\mu$-type or $y$-type distortion. The energy deposition, $Q(t)$, could be fast or small but extended in time; the integral is comovil, simply accumulates this energy over time weighted by the $a^4$ which is in the dependence of the photon energy density.

The visibility functions can be approximated as~\cite{Chluba:2016bvg}
\begin{align}
\mathcal{J}_{\mu}(z) &\simeq
\exp[-(\tfrac{z}{z_{\mathrm{th}}})^{5/2}]\,
\big[1-\exp(-(\tfrac{1+z}{1+z_{\mu y}})^{1.88})\big],\\
\mathcal{J}_{y}(z) &\simeq
[1+(\tfrac{1+z}{6\times10^{4}})^{2.58}]^{-1} \,,
\end{align}
where $z_{\mu y} \approx 5.8 \times 10^{4}$ marks the transition between the $\mu$– and $y$–eras, and  
$z_{\mathrm{th}} \approx 1.98 \times 10^{6}$ is the thermalization redshift \cite{Hu:1992dc}, above which distortions are completely erased. Throughout this work, we assume a sharp boundary between the $\mu$ and $y$ regimes at $z_{\mu y}$.

Finally, we calculate the SDs amplitudes generated by the various dark sector decay channels and injection histories discussed in the previous section.  This allows us to connect the microphysical decay parameters, such as the lifetime $\Gamma_\chi^{-1}$, the confinement scale $a_c$, and the relativistic velocity dispersion $v_c$, with observable $\mu$ and $y$ distortions in the CMB spectrum.

%%%%%%% ----------------- Gamma -- Sigma -------------------
\begin{figure*}[t]
    \begin{center}
    \includegraphics[width=0.32\textwidth]{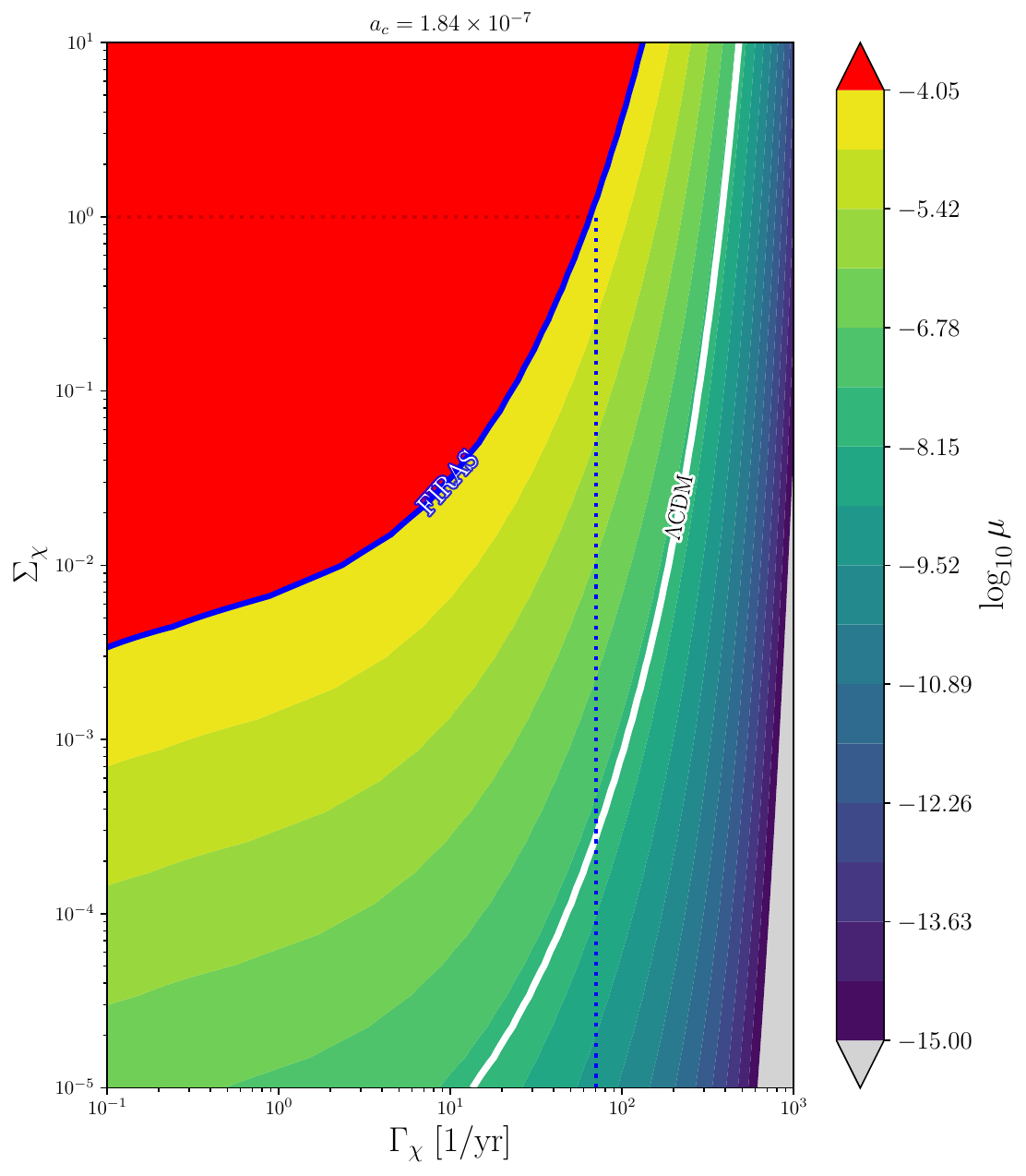}
    \includegraphics[width=0.32\textwidth]{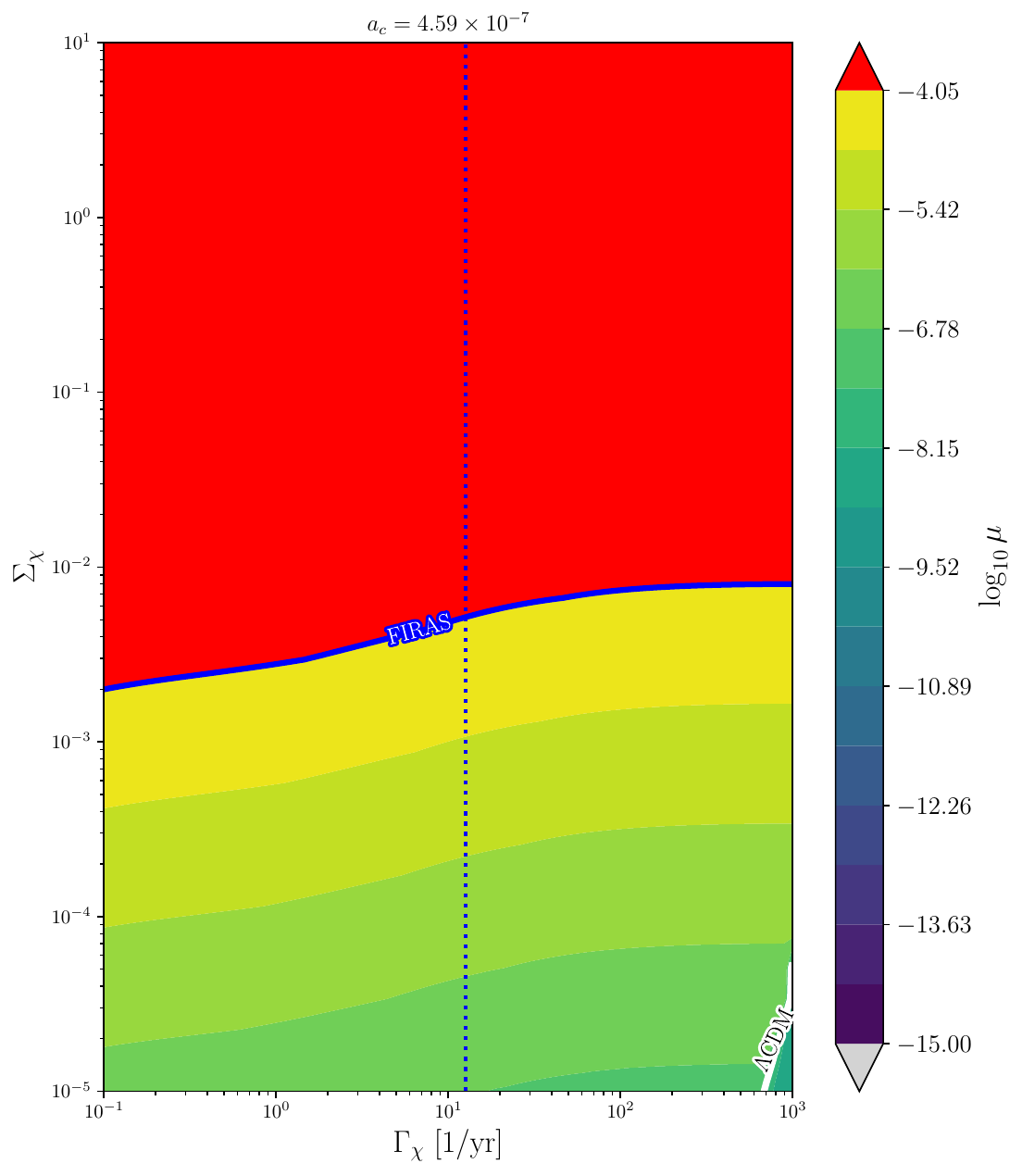}
    \includegraphics[width=0.32\textwidth]{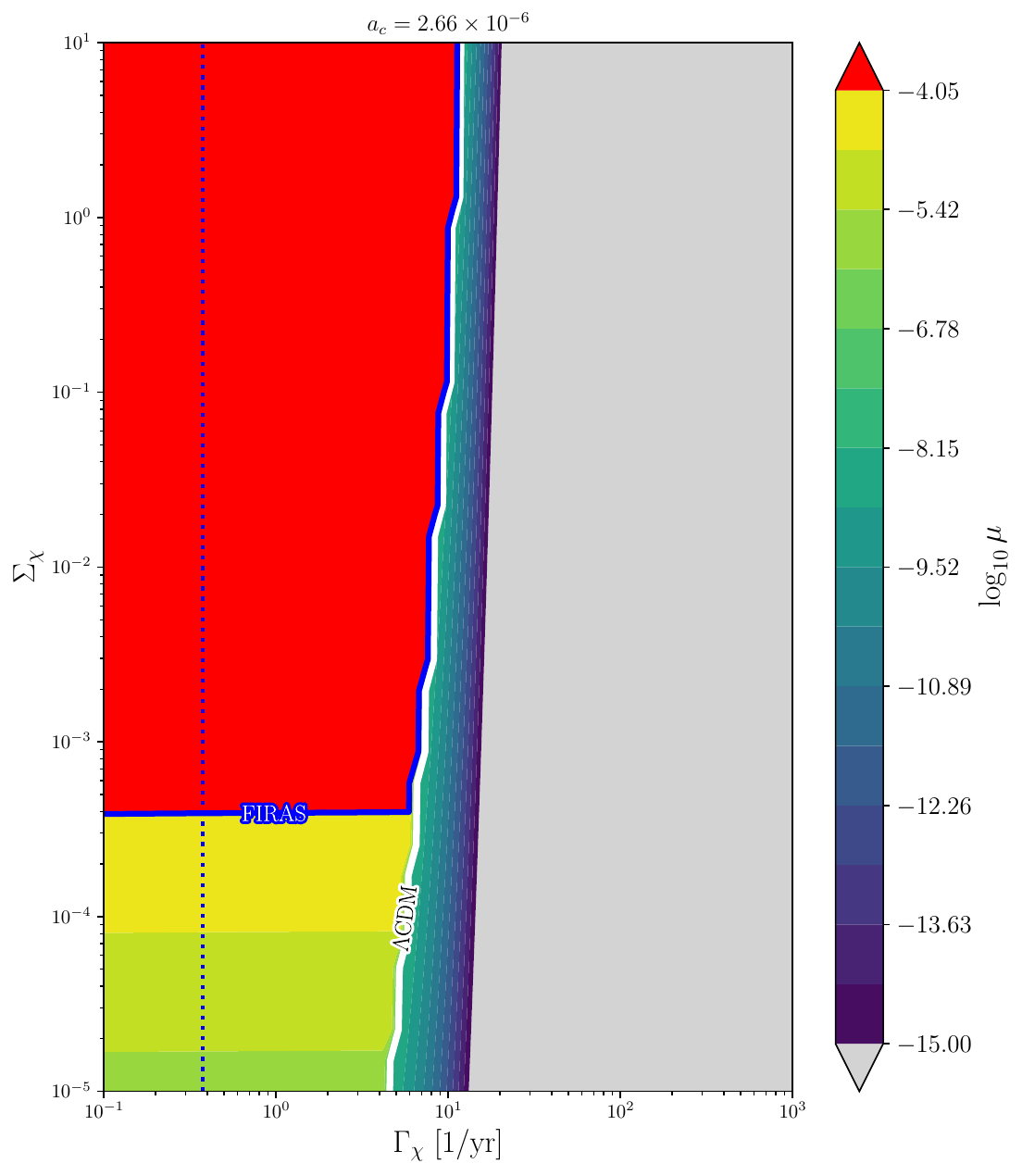}
    \caption{Contour plots illustrating the magnitude of the $\mu$-type SD in the parameter space defined by the decay rate and the fractional abundance parameter ($\Gamma_{\chi}-\Sigma_{\chi}$) for three fixed values of the transition scale factor $a_c$: upper bound for $\mu_{\rm firas}$ constraint (left panel), upper CMB limit with $v_{dmc} = 1/\sqrt{2}$ (middle panel), and upper CMB limit with $v_{dmc} = 0$ (right panel). The solid white line labeled $\Lambda$CDM corresponds to the $\mu$-SD value predicted by the standard cosmological model. The blue isoline labeled FIRAS represents the observational upper limit from FIRAS. Regions above FIRAS isoline are excluded (red-shaded areas). The grey-shaded areas represent parameter regions producing produce negligible distortions (undetectable or at/below $\Lambda$CDM). Vertical dashed lines mark the critical decay rate $\Gamma_{cr}(a_c)$ distinguishing short- from long-lived regimes.}
    \label{fig:gamma_vs_sigma}
    \end{center}
\end{figure*}

% ---------------------------- %%
\section{Results}\label{sec_4} %%
% ---------------------------- %%
We aim to determine the observational signatures of $\mu$- and $y$-SD in the previously discussed QCD-DM scenario. We computed the SD magnitudes for benchmark parameters and also derived constraints in parameter space using observational upper bounds. The parameter space considered includes the scale factor at the QCD-DM transition ($a_{c}$), the initial velocities of the decaying particles, $v_{\chi c}$, the parameter $\Sigma_{\chi} = f_{\rm eff} f_{\chi}$ quantifying the effective fraction of the energy going into CMB photons from the decaying species, and the decay parameters of the different channels, for instance, $\Gamma_{\chi}$ for the exponential, $\gamma$ and $\alpha$ for the power-law. The velocity dispersion of the stable dark matter, $v_{\rm dmc}$, influences only the Hubble parameter $H$, resulting in sub-percent differences in the computed spectral distortions; therefore, it is not considered a relevant parameter in our analysis.

Benchmark QCD-DM values were taken from \cite{Mastache:2019bxu, Torres:2025qko}, where the authors analyze the impact of the Bound Dark Matter model, which is a particular QCD--DM model, using CMB Planck 2018 data, Baryon Acoustic Oscillations (BAO), Supernovae Ia (JLA) catalogs, and DESI BAO measurements. Monte Carlo simulations constrain the parameter space $a_c - v_c$, obtaining the upper limits for $a_{c}$. Specifically, for an abrupt transition ($v_{dmc} = 0$), they find $a_c \leq 2.66 \times 10^{-6}$ at $1\sigma$ and $a_c \leq 5.62 \times 10^{-6}$ at $2\sigma$ C.L.; while for a smoother transition ($v_{dmc} = 1/\sqrt{2}$), the limits are $a_c \leq 4.59 \times 10^{-7}$ at $1\sigma$ and $a_c \leq 7.19 \times 10^{-7}$ at $2\sigma$ C.L. Interestingly, the upper bound for the transition is close to the redshift at which $\mu$-SD started being quantified, $z_\mu \simeq 4 \times 10^6$. In \cite{Mastache:2019bxu}, it is shown that QCD-DM can mimic the small-scale suppression effects of WDM, requiring a value of $a_c = 3.18\times 10^{-7}$ with $v_{dmc} = 0$. This combination of parameters yields the same free-streaming scale and suppression in the matter power spectrum as a thermal 3 KeV WDM particle.

For observational constraints, we consider measurements from FIRAS, with values of $\mu_{\rm firas} < 9\times10^{-5}$ and $y_{\rm firas}<1.5\times10^{-5}$ \cite{Fixsen:1996nj}. For reference, we use $\mu_{\Lambda \rm cdm}=2\times10^{-8}$ and $y_{\rm \Lambda cdm} =4\times 10^{-9}$ as the canonical $\Lambda$CDM prediction \cite{Chluba:2016bvg}. The expected $y$-SD, including late-time effects, for instance, thermal Sunyaev–Zeldovich effect, reionization, or early structure formation shocks, is taken as $y_{\rm sz} = 1.7 \times 10^{-6}$ \cite{Hill:2015tqa, Rotti:2020rdl, Chluba:2019nxa}.

\subsection{Oscillatory and Cascade Decays: Exponential Approximations}
Figure~\ref{fig:caseB} illustrates the heating rate evolution $\dot Q/(H\rho_\gamma)$ for different decay mechanisms. First, let us focus on the oscillatory and a two-step (cascade) decay, for which we compared them with their respective exponential approximations. In both cases, the exponential mapping introduced in Sec.~\ref{subsec:energy_deposition} reproduces the energy injection of both decays.

Quantitatively, Table~\ref{tab:decay_osc_2step} summarizes the comparison between the exact and approximate (exponential) computations of the $\mu$- and $y$-type spectral distortions for both oscillatory and cascade decays. For the oscillatory decay, the ratios $(\mu_{\rm exact}/\mu_{\rm approx},\, y_{\rm exact}/y_{\rm approx})$ are close to unity, with deviations below $2\%$ in $\mu$ and at the sub-percent level in $y$, confirming that the exponential mapping accurately reproduces the energy release even for highly oscillatory cases.  For the cascade decay, larger deviations appear in $\mu$, up to a factor of $\sim3$ for slow-decaying intermediates, while the $y$ distortion remains within $\sim30\%$. These differences are consistent with the delayed onset of photon injection in two-step decays, which slightly shifts the peak of the effective heating rate but leaves the overall distortion amplitude of the same order of magnitude.  Overall, the ratios in Table~\ref{tab:decay_osc_2step} demonstrate that both complex decay modes can be effectively reduced to an exponential description with minor quantitative corrections, preserving the phenomenological impact on CMB spectral distortions within observational accuracy.

Given these results, the subsequent analysis focuses on the parameter space of the exponential and power-law decay models, since both oscillatory and two-step (cascade) decay processes can be consistently reinterpreted within the exponential framework through effective redefinitions of the decay rate and normalization parameters. This simplification allows a unified treatment of diverse decay mechanisms using a minimal set of parameters that govern the energy injection history.

\subsection{Parameter space for exponential decay}

%%%% ------------- ac - Sigma ---------------%%%%
\begin{figure}[t]
\begin{center}
    \includegraphics[width=0.49\textwidth]{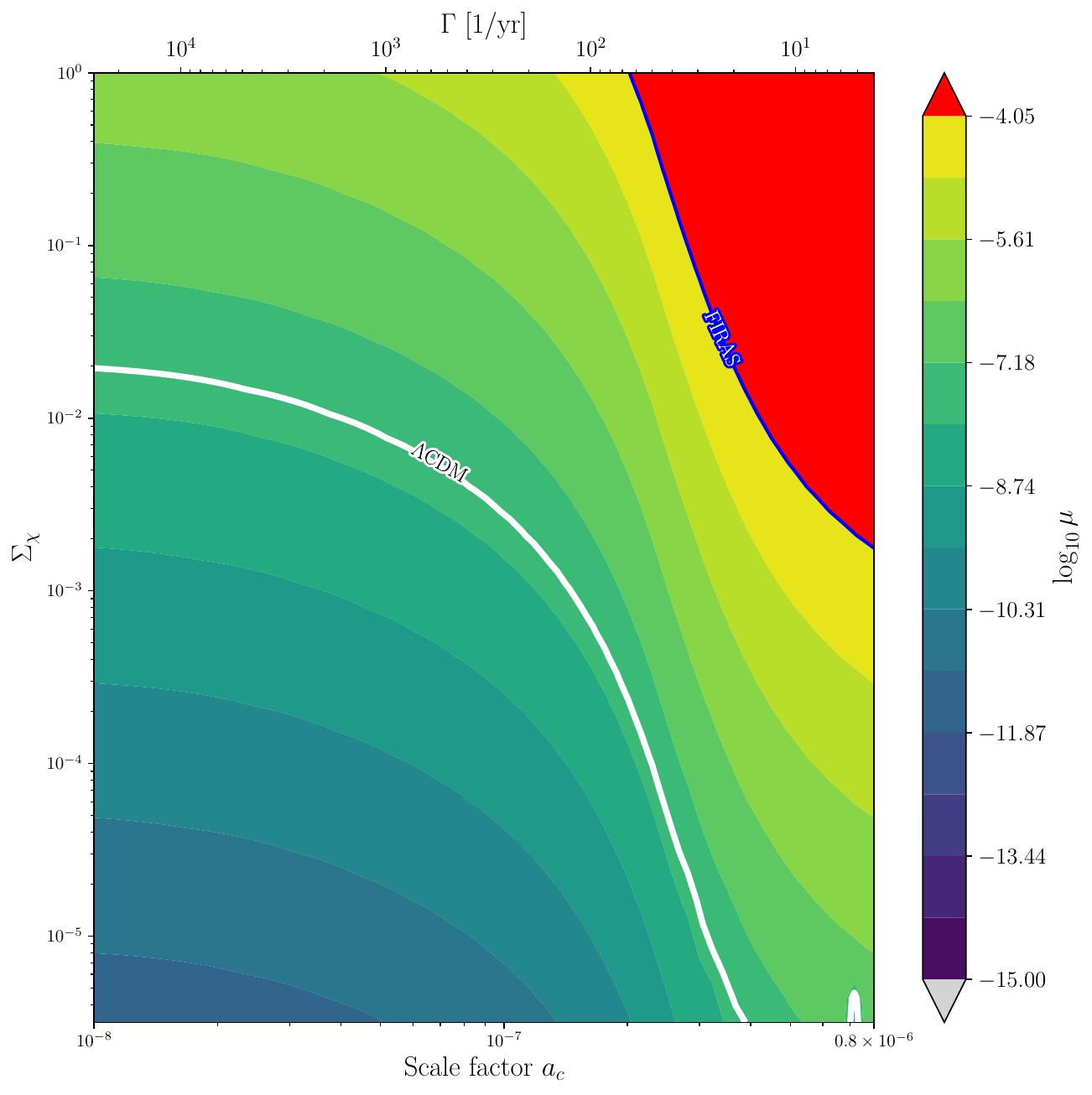}    
    \caption{Contour plot of the $\mu$-type SD in the parameter space ($a_c$-$\Sigma_\chi$) for $\Gamma_\chi = 4H_c$. The solid white isoline represents the canonical value that $\Lambda$CDM predicted. The blue isoline marks the observational upper limit obtained from FIRAS, the red-shaded region is excluded by observations.}
    \label{fig:sigma_vs_ac_firas}
\end{center}
\end{figure}
%%%% ------------- ac - Sigma ---------------%%%%

%%%%%% ---------------------- Sigma - Gamma ------------------------
\begin{figure}[t!]
\centering
    \includegraphics[width=\columnwidth]{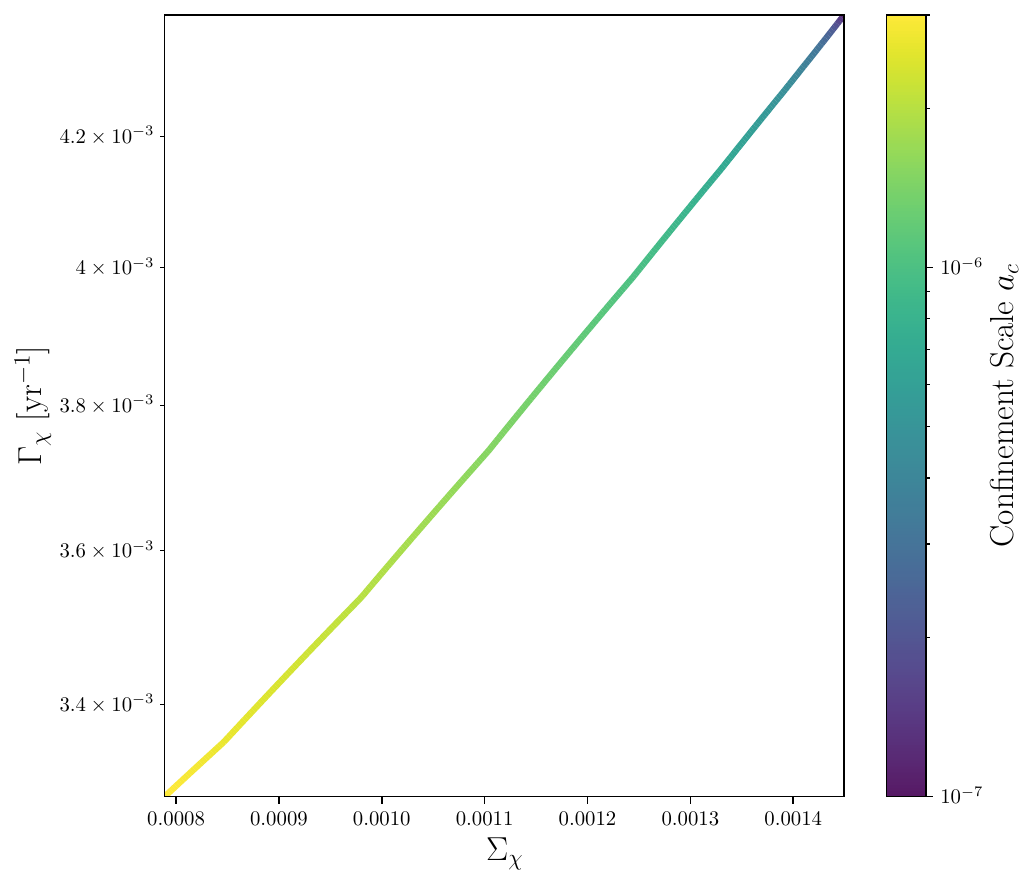}
    \caption{
    Constraints on the parameter space $(\Sigma_\chi,\,\Gamma_\chi)$ for the exponential decay model at fixed $v_{\chi c}\simeq1$. 
    The curve corresponds to the combination of $(\Sigma_\chi,\,\Gamma_\chi)$ that, for a given confinement scale $a_c$ (indicated by color), 
    simultaneously reproduces the FIRAS observational bounds on the spectral distortions $\mu=\mu_{\mathrm{firas}}$ and $y=y_{\mathrm{firas}}$. 
    The color gradient thus traces the dependence of the allowed parameter space on $a_c$.}
    \label{fig:gamma_sigma_ac}
\end{figure}
%%%%%% ---------------------- Sigma - Gamma ------------------------

We constrain the parameter $\Sigma_{\chi}$ by assuming that at the moment of the confinement, the decaying particle has a rate $\Gamma_{\chi}=4H_c$, representing an intermediate regime between instantaneous and slow decay. We explore the parameter space of the QCD–DM model under the assumption that the decay constant at the moment of transition satisfies $\Gamma_\chi = 4H_c$, representing an intermediate regime between instantaneous and slow decay. However, in the Appendix \ref{app:decaying_rate} we show that the amplitude of the spectral distortion increases approximately linearly with the decay rate $\Gamma_\chi$. 

Within this framework, $\Sigma_\chi$ is the energy–injection fraction in the model, but in Fig.~\ref{fig:sigma_x_constraints} it also plays the role of a relative normalization referenced to the FIRAS limit. The $\Sigma_\chi$ helps to rescale the total $\mu-$ or $y-$type SD, for any chosen $(a_c, v_{\chi c})$, the color is the value that reproduces the target spectral–distortion amplitude $(\mu=\mu_{\rm firas}$ or $y=y_{\rm firas}$). Therefore, $\log_{10}\Sigma_\chi<0$ indicates that the un–rescaled model would produce distortions above the FIRAS bound (and must be scaled down, $\Sigma_\chi<1$), whereas $\log_{10}\Sigma_\chi>0$ signals that the un–rescaled model would naturally yield distortions below the FIRAS limit (requiring $\Sigma_\chi>1$ to reach the target value). Positive values are therefore not observationally excluded; instead, they quantify how far the model sits below observational bounds and are useful for mapping the scaling behavior across parameter space.

The cyan vertical line in Fig.~\ref{fig:sigma_x_constraints} corresponds to the $1\sigma$ CMB upper bound on the confinement scale, $a_c = 2.66\times10^{-6}$, with $\Gamma_\chi = 0.38~{\rm yr^{-1}}$. At this transition, the $\mu$-type distortion imposes a stronger constraint than the $y$-type, requiring $\log_{10}\Sigma_\chi \lesssim -2.37$. 

%%%%%%%-------------- EXPONENTIAL MODEL ---------_%%%%%
\begin{table*}[t]
\centering
\scriptsize
\setlength{\tabcolsep}{3pt}
\renewcommand{\arraystretch}{1.15}
\begin{tabular}{lccccc}
Case & Target value & $a_c$ & $\log_{10}\Sigma_\chi$ & $\Gamma_\chi$ [1/yr] & Associated SD$^{\,a}$ ($v_{\chi c}\!\sim\!0\,/\,v_{\chi c}\!\sim\!1$) \\ 
\hline\hline
\multicolumn{6}{c}{\bf y–type spectral distortion} \\
\hline
CMB & $a_c=2.66\times10^{-6}$ & --- & $[-1.50,-0.33]$ & 0.38 & $1.02\!\times\!10^{-2}$ / $1.25\!\times\!10^{-2}$ \\[2pt]
FIRAS & $y_{\rm firas}=1.5\times10^{-5}$ & $[1.92,2.42]\!\times\!10^{-6}$ & $0.0$ & [0.72,0.45] & $1.96\!\times\!10^{-2}$ / $2.85\!\times\!10^{-1}$ \\[2pt]
$\Lambda$CDM & $y_{\rm\Lambda cdm}=4.0\times10^{-9}$ & $[1.29,1.40]\!\times\!10^{-6}$ & $3.57$ & [1.59,1.36] & $4.10\!\times\!10$ / $6.91\times10^2$ \\
SZ & $y_{\rm sz}=1.7\times10^{-6}$ & $[1.67,1.96]\!\times\!10^{-6}$ & $0.95$ & [0.95,0.69] & $1.39\!\times\!10^{-1}$ / $2.16$ \\[2pt]
\hline\hline
\multicolumn{6}{c}{\bf $\mu$–type spectral distortion} \\
\hline
CMB & $a_c=2.66\times10^{-6}$ & --- & [$-3.63,-2.37$] & 0.38 & $3.23\!\times\!10^{-11}$ / $ 2.66\!\times\!10^{-11}$ \\
FIRAS & $\mu_{\rm firas}=9.0\times10^{-5}$ & $[1.84,2.49]\!\times\!10^{-7}$ & $0.0$ & [78.4,42.7] & $<10^{-15}$ / $<10^{-15}$ \\[2pt]
$\Lambda$CDM & $\mu_{\rm\Lambda cdm}=2.0\times10^{-8}$ & $[1.13,1.22]\!\times\!10^{-7}$ & 3.65 & [206.7,178.0] & $<10^{-15}$ / $<10^{-15}$ \\[2pt]
3keV WDM & $a_c=3.2\times10^{-7}$ & --- & [$-1.88,-0.66$] & 26.19 & $<10^{-15}$ / $<10^{-15}$ \\[2pt]
\end{tabular}
\caption{Benchmark cases for the QCD--DM model and their corresponding $\mu$– and $y$–type spectral distortions. Each row lists representative values of $a_c$, $\log_{10}\Sigma_\chi$, and $\Gamma_\chi$ producing the associated distortion, $\mu-$sd when targeting $y$-sd and vice-versa. CMB and FIRAS cases correspond to observational upper limits, $\Lambda$CDM reproduces the standard model prediction, WDM mimics a 3~keV warm dark matter scenario, and SZ represents late-time $y$ distortions. Ranges are ordered for the cases $[\,v_{\chi c}\!\sim\!1, v_{\chi c}\!\sim\!0\,]$. Values out of the observable sensitivity are displayed as $<10^{-15}$.}
\label{tab:compact_mu_y}
\end{table*}
%%%%%%%-------------- EXPONENTIAL MODEL ---------_%%%%%

%%%%%%   ------------ BEST FIT EXPONENTIAL ---------%%
\begin{table}[h]
\centering
\small
\setlength{\tabcolsep}{3pt}
\renewcommand{\arraystretch}{1.1}
\begin{tabular}{lcc}
\hline
Parameter & $v_{\chi c}\!\simeq\!0.99$ & $v_{\chi c}\!\simeq\!0.001$ \\ 
\hline
$a_c$ & \multicolumn{2}{c}{$2.66\times10^{-6}$} \\[2pt]
$\Sigma_\chi$ & $8.47\times10^{-4}$ & $3.20\times10^{-7}$ \\[2pt]
$\Gamma_\chi$ [yr$^{-1}$] & $3.35\times10^{-3}$ & $4.00\times10^{-3}$ \\[2pt]
\hline
\end{tabular} 
\caption{Best-fit parameters at fixed $a_c$ given the CMB constraints for $a_c$ for the two extreme values for the velocity $v_{\chi c}$ for which we simultaniusly obtain $\mu = \mu_{\rm firas}$ and $y = y_{\rm firas}$.}
\label{tab:bestfit_fixed_ac}
\end{table}
%%%%%%   ------------ BEST FIT EXPONENTIAL ---------%%

The red isoline marks the FIRAS limit, corresponding to the observational bounds $\mu=\mu_{\rm firas}=9\times10^{-5}$ and $y=y_{\rm firas}=1.5\times10^{-5}$. These values define the largest confinement scale allowed by current spectral-distortion measurements for $\log_{10}\Sigma_\chi=0$. The corresponding confinement scales occurs at $a_c\simeq(1.84\!-\!2.49)\times10^{-7}$, near the onset of the $\mu$-era, with decay rates in the range $\Gamma_\chi=(78.4\!-\!42.7)\,{\rm yr^{-1}}$. Interestingly, even with high decay rates, $\mu$-type SD remain below observable thresholds indicating early energy injection is efficiently thermalized.

The white isoline in Fig.~\ref{fig:sigma_x_constraints} indicates the primordial $\Lambda$CDM distortion level ($\mu_{\Lambda{\rm cdm}}=2\times10^{-8}$, $y_{\Lambda{\rm cdm}}=4\times10^{-9}$). This case corresponds to a confinement scale at $a_c=(1.13\!-\!1.22)\times10^{-7}$ and $\Gamma_\chi=(206.7\!-\!178.0)\,{\rm yr^{-1}}$, with $\log_{10}\Sigma_\chi = \log_{10}\mu_{\rm firas}/\mu_{\rm \Lambda cdm} \simeq3.65$. Smaller confinement scales would produce distortions indistinguishable from the standard model. Notably, this result shows that the $\mu$-type constraint is generally more stringent than the $y$-type, and that spectral distortions can impose stronger limits on QCD–DM scenarios than the CMB power spectrum alone. The model reproduces standard predictions with high decay rates and very early $a_c$, confirming QCD–DM can recover standard cosmology in limiting cases.

The magenta isoline marks the transition scale that mimics the matter power-spectrum cutoff of a 3~keV warm dark matter (WDM) model, occurring at $a_c=3.2\times10^{-7}$ and $\Gamma_\chi=26.2~{\rm yr^{-1}}$. In this case, the effective energy fraction satisfies $\log_{10}\Sigma_\chi\lesssim -0.66$. Results in unobservable $y$-SD. The brown isoline, by contrast, corresponds to the late-time astrophysical $y$-distortion reference, $y=y_{\rm sz}=1.7\times10^{-6}$, fixing the confinement scale at $a_c=(1.67\!-\!1.96)\times10^{-6}$.

%\textbf{Constraints in the $\Gamma_\chi$–$\Sigma_\chi$ plane.}
The parameter space spanned by $(\Gamma_\chi,\Sigma_\chi)$ for fixed $a_c$ is shown in Fig.~\ref{fig:gamma_vs_sigma}. At the CMB upper limit $a_c=2.66\times10^{-6}$ (right panel), with $v_{\chi c}\sim 1$, the blue dashed line denotes $\Gamma_\chi=4H_c=0.38~{\rm yr^{-1}}$, separating long- and short-lived particles. For fast decays, the allowed region is bounded by $\Sigma_\chi\leq3.7\times10^{-4}$ and $\Gamma_\chi\lesssim6.5~{\rm yr^{-1}}$; faster decays produce negligible or indistinguishable distortions. At intermediate confinement scales ($a_c=4.59\times10^{-7}$, middle panel), the blue vertical line represent $\Gamma_\chi=4H_c=12.7~{\rm yr^{-1}}$, and the upper limit for detectable distortions is $\Sigma_\chi\simeq8\times10^{-3}$.  
For earlier transitions ($a_c=1.84\times10^{-7}$, left panel) $\Gamma_\chi = 4H_c = 78.4~{\rm yr^{-1}}$, and can go up to $\Sigma_\chi=1$ for $\mu=\mu_{\rm firas}$, a nonlinear correlation emerges between $\Gamma_\chi$ and $\Sigma_\chi$: higher decay rates require smaller $\Sigma_\chi$ to maintain the same $\mu$-amplitude. For example, $\Gamma_\chi=380~{\rm yr^{-1}}$ with $\Sigma_\chi=1$ yields $\mu=\mu_{\Lambda{\rm cdm}}$, or equivalently, $\Gamma_\chi=4H_c$ with $\Sigma_\chi=2.8\times10^{-4}$.

% Earlier confinement epochs (smaller a_c) 
% later transitions (larger a_c) 
The range values for each of the benchmark case (always assuming $\Gamma_\chi = 4 H_c$) are shown in Table~\ref{tab:compact_mu_y} for both $\mu$- and $y$- type SD where one can conclude that earlier confinement epochs get the value of $y_{\rm firas}$ SD with amplitudes for $\mu$-SD larger than $\mu_{\rm firas}$; later transitions produce $\mu_{\rm firas}$ distortions but with negligible $y$- amplitudes. Therefore, the decay rate $\Gamma_\chi$ significantly affects whether distortions are above or below detection thresholds (FIRAS sensitivity or future PIXIE-like experiments). Relativistic decays (with $v_{\chi c} \sim 1$) tend to generate larger spectral distortions, as particles inject more energy into the photon bath.

There is a degeneracy between $a_c$ and $\Gamma_\chi$: slower decays can be compensated by larger $\Sigma_\chi$, as reflected in the $\Lambda$CDM isoline of Fig.~\ref{fig:sigma_vs_ac_firas}. There exists a critical scale factor below which the distortion signal becomes negligible, $\mu(a_c)<\mathcal{O}(10^{-15})$. This threshold can be estimated from the condition $\Gamma_\chi (t-t_c)\simeq 1$, yielding
\begin{equation}
    a_{\rm crit}^2 = a_{\rm firas}^2\,\frac{1}{4H_0t_0(\delta^2-1)},
\end{equation}
where $\delta$ quantifies the relative shift between $a_c$ and the effective critical scale $a_{\rm crit}$. Taking $\delta=1.04$, the critical value for the FIRAS case is $a_{\rm crit}\simeq9.1\times10^{-7}$, below which the $\mu$- and $y$- type signals are effectively erased.

Overall, individual FIRAS observations for $\mu$- and $y$- SDs exclude large regions of the $(\Gamma_\chi,\Sigma_\chi)$ plane, particularly for later transitions, high energy transfer, and slow decaying particles. The $\mu$-type constraint remains the most restrictive observable, providing limits roughly an order of magnitude tighter than $y$-type distortions. CMB data delimit the high-$a_c$ regime, while spectral distortions dominate for earlier transitions ($a_c\lesssim10^{-6}$). Together, these complementary bounds delineate the viable parameter space for exponential decays in the QCD–DM scenario.

We have examined the constraints on the parameter space of the QCD--DM scenario using either $\mu$- or $y$-type SDs independently. However, a more robust constraint emerges when both observational bounds from FIRAS, $\mu_{\rm firas}$ and $y_{\rm firas}$, are simultaneously satisfied. Figure~\ref{fig:gamma_sigma_ac} shows the resulting allowed region in the $(\Sigma_\chi,\,\Gamma_\chi)$ parameter space for the exponential decay model at fixed initial velocity $v_{\chi c} \simeq 1$. The curve corresponds to the combination of parameters $(\Sigma_\chi,\,\Gamma_\chi)$ that, for a given confinement scale $a_c$ (indicated by the color gradient), reproduces both FIRAS limits. 

The analysis reveals that the joint constraint tightly localizes the decay rate around $\Gamma_\chi \simeq (3.3 - 4.4) \times10^{-3}\,{\rm yr^{-1}}$, indicating that the parameter space for the particle lifetime is narrowly defined once both $\mu$ and $y$ are simultaneously fitted and $v_{\chi c}\to 1$. The interaction strength $\Sigma_\chi$ exhibits a monotonic decrease as the confinement epoch $a_c$ increases: for early transitions, such as $a_c = 1.0 \times 10^{-7}$, the required coupling is $\Sigma_\chi = 1.45 \times 10^{-3}$, whereas for later transitions approaching the CMB upper limit ($a_c = 2.66 \times 10^{-6}$), it decreases to $\Sigma_\chi = 8.47 \times 10^{-4}$. This trend reflects a compensation mechanism in which a stronger coupling at early epochs offsets the reduced efficiency of energy transfer to the photon bath, while at later epochs, smaller $\Sigma_\chi$ values suffice as the decay occurs closer to the thermalization window of maximal SD production.

Now, we compare the case when $v_{\chi c}\to 0$, see values in Table~\ref{tab:bestfit_fixed_ac} the best decay rate of $\Gamma_\chi = 4 \times 10^{-3}~{\rm yr^{-1}}$ for the CMB upper limit, significantly below the critical decay threshold $4H(a_c) \simeq 0.38~{\rm yr^{-1}}$ that separates fast from slow decays. This places the preferred solution in the slow-decay regime, where the energy deposition into the photon bath is gradual and the resulting distortions are sensitive to both the temporal profile of the decay and the coupling strength.

The corresponding interaction strength, $\Sigma_\chi = 3.2 \times 10^{-7}$, lies well below the upper limit allowed for fast decays, $\Sigma_\chi \leq 3.7 \times 10^{-4}$. This emphasizes that, for slow decays, a smaller coupling suffices to saturate the FIRAS bounds without exceeding them, due to the prolonged duration of energy injection. On the other hand, decay rates larger than $\Gamma_\chi \gtrsim 6.5~{\rm yr^{-1}}$ lead to rapid energy release that either thermalizes before contributing to distortions or produces effects indistinguishable from standard scenarios, thus becoming observationally irrelevant.

Altogether, these results highlight that the joint $\mu$--$y$ constraint defines a region in $(\Sigma_\chi,\,\Gamma_\chi)$ space: decay rates are preferred to be slow to moderate, $\Gamma_\chi \lesssim 6.5~{\rm yr^{-1}}$, and interaction strengths limited to $\Sigma_\chi \lesssim 3.7 \times 10^{-4}$, with best-fit solutions clustering around $\Gamma_\chi \sim 10^{-3}~{\rm yr^{-1}}$ and $\Sigma_\chi \sim 10^{-7}$.

The figure ~\ref{fig:gamma_sigma_ac} also demonstrates that the degeneracy between $\Sigma_\chi$ and $\Gamma_\chi$ could be broken when both $\mu$-- and $y$--type distortions are taken into account: while $\Gamma_\chi$ remains nearly constant across all fitted cases, $\Sigma_\chi$ evolves predictably with $a_c$. Altogether, these results highlight how combining $\mu$ and $y$ spectral distortion data offers a powerful and complementary probe of the QCD--DM phenomenology, effectively linking the confinement scale $a_c$, decay rate $\Gamma_\chi$, and energy efficiency $\Sigma_\chi$ into a narrow, observationally consistent parameter region.

\subsection{Parameter space for power-law decay}

%%%%%%% ---------------  POWER-LAW VALUES --------------- %%%
\begin{table*}[t]
\centering
\scriptsize
\setlength{\tabcolsep}{4pt}
\renewcommand{\arraystretch}{1.15}
\begin{tabular}{llcccc}
\hline\hline
Case & Target & $a_c$ range & $\log_{10}\Sigma_\chi$ & $\gamma_\chi$ [yr$^{-1}$] & Associated SD ($v_{\chi c}\!\sim\!0\;/\;v_{\chi c}\!\sim\!1$)\\
\hline\hline
\multicolumn{6}{c}{\bf $\alpha=0.5$}\\
\hline
\multirow{2}{*}{$\Lambda$CDM}
& $\mu_{\Lambda\rm cdm}=2.0\times10^{-8}$ & $[1.22,1.41]\!\times\!10^{-7}$ & \multirow{2}{*}{$-3.65$} & $[176.60,133.05]$ & $y < 10^{-15}/10^{-15}$\\
& $y_{\Lambda\rm cdm}=4.0\times10^{-9}$     & $[1.42,1.56]\!\times\!10^{-6}$ & & $[1.32,1.09]$   & $\mu=3.48\times10^{-3}/6.07\times10^{-2}$\\[2pt]
\multirow{2}{*}{FIRAS}
& $\mu_{\rm firas}=9.0\times10^{-5}$ & $[2.21,3.17]\!\times\!10^{-7}$ & \multirow{2}{*}{$0.00$} & $[53.96,26.32]$ & $y < 10^{-15}/10^{-15}$\\
& $y_{\rm firas}=1.5\times10^{-5}$   & $[2.27,3.00]\!\times\!10^{-6}$ &  & $[0.52,0.30]$   & $\mu=6.90\times10^{-3}/1.01\times10^{-1}$\\[2pt]
\multirow{2}{*}{3 KeV WDM}
& \multirow{2}{*}{$a_c=3.18\times10^{-7}$} & --- & $[-1.26,-0.01]$ (at $\mu$) & \multirow{2}{*}{$26.19$} & $y < 10^{-15}/10^{-15}$\\
& & --- & $[91.32,91.70]$ (at $y$)  &  & $\mu=9.12\times10^{-5}/1.64\times10^{-3}$\\[2pt]
\multirow{2}{*}{CMB}
& \multirow{2}{*}{$a_c=2.7\times10^{-6}$} & — & $[-3.12,-1.83]$ & \multirow{2}{*}{$0.38$} & $y=5.31\times10^{-6}/8.37\times10^{-5}$\\
& & — & $[-0.75,0.45]$  & & $\mu=6.11\times10^{-3}/1.19\times10^{-1}$\\[2pt]
SZ & $y_{\rm sz}=1.7\times10^{-6}$ & $[1.92,2.36]\!\times\!10^{-6}$ & $0.95$ & $[0.72,0.48]$ & $\mu=5.39\times10^{-3}/8.45\times10^{-2}$\\
\hline\hline
\multicolumn{6}{c}{\bf $\alpha=6$}\\
\hline
\multirow{2}{*}{$\Lambda$CDM}
& $\mu_{\Lambda\rm cdm}=2.0\times10^{-8}$ & $[1.10,1.22]\!\times\!10^{-7}$ & \multirow{2}{*}{$-3.65$} & $[219.33,178.81]$ & $y=0/0$\\
& $y_{\Lambda\rm cdm}=4.0\times10^{-9}$     & $[1.42,1.56]\!\times\!10^{-6}$ & & $[1.32,1.09]$    & $\mu=4.17\times10^{-2}/7.28\times10^{-1}$\\[2pt]
\multirow{2}{*}{FIRAS}
& $\mu_{\rm firas}=9.0\times10^{-5}$ & $[1.78,2.28]\!\times\!10^{-7}$ & \multirow{2}{*}{$0.00$} & $[83.98,51.01]$ & $y=0/0$\\
& $y_{\rm firas}=1.5\times10^{-5}$   & $[2.27,3.00]\!\times\!10^{-6}$ &  & $[0.52,0.30]$   & $\mu=8.28\times10^{-2}/1.21$\\[2pt]
\multirow{2}{*}{3 KeV WDM}
& \multirow{2}{*}{$a_c=3.18\times10^{-7}$} & --- & $[-2.34,-1.08]$ (at $\mu$) & \multirow{2}{*}{$26.19$} & $y < 10^{-15}/10^{-15}$\\
&  & --- & $[91.32,91.70]$ (at $y$)  &  & $\mu=1.09\times10^{-3}/1.96\times10^{-2}$\\[2pt]
\multirow{2}{*}{CMB}
& \multirow{2}{*}{$a_c=2.7\times10^{-6}$} & — & $[-4.20,-2.91]$ & \multirow{2}{*}{$0.38$} & $y=5.31\times10^{-6}/8.37\times10^{-5}$\\
& & — & $[-0.75,0.45]$  & & $\mu=7.33\times10^{-2}/1.43$\\[2pt]
SZ & $y_{\rm sz}=1.7\times10^{-6}$ & $[1.92,2.36]\!\times\!10^{-6}$ & $0.95$ & $[0.72,0.48]$ & $\mu=6.46\times10^{-2}/1.01$\\
\hline\hline
\end{tabular}
\caption{Summary of benchmark constraints for power–law energy injection with $\alpha=0.5$ and $\alpha=6$.  
Each block collates, for the indicated target ($\mu$ or $y$), the corresponding $(a_c,\log_{10}\Sigma_\chi,\Gamma_\chi)$ ranges and the \emph{associated} complementary distortion (last column) at $v_{\chi c}\!\sim\!0$ and $v_{\chi c}\!\sim\!1$.  
These results are intended to be read jointly with the $(a_c,v_{\chi c})$ contour plots (color–coded in $\mu$ and $y$) for the same $\alpha$ values.}
\label{tab:powerlaw_summary}
\end{table*}
%%%%%%% ---------------  POWER-LAW VALUES --------------- %%%

For the power-law decay model, the confinement scale $a_c$, decay rate $\Gamma_\chi$, and effective energy deposition $\Sigma_\chi$ were determined for representative benchmark cases listed in Table~\ref{tab:powerlaw_summary}. 

For the case with $\alpha = 0.5$, the resulting spectral distortions follow the expected trend of smoother and broader energy release. The FIRAS and CMB bounds appear at similar confinement scales to those found in the exponential decay scenario, with $a_c\simeq(2.2$–$3.0)\times10^{-7}$ for $\mu_{\rm firas}$ and $a_c\simeq(2.3$–$3.0)\times10^{-6}$ for $y_{\rm firas}$. The corresponding decay rates range from $\Gamma_\chi\simeq(54$–$26)\,{\rm yr^{-1}}$ in the $\mu$-era to $(0.5$–$0.3)\,{\rm yr^{-1}}$ in the $y$-era. At these scales, the induced $\mu$- and $y$-type distortions increase with $v_{\chi c}$, reaching an order of magnitude higher for relativistic decays ($v_{\chi c}\!\sim\!1$). The $\Lambda$CDM-like transitions at $a_c\!\sim\!10^{-7}$ reproduce the expected small distortions, while the $3\,{\rm keV}$ WDM analog ($a_c\!=\!3.2\times10^{-7}$) yields negligible $y$-type distortion but a moderate $\mu$-type signal, consistent with early-time energy release. 

For the case with $\alpha = 6$, the energy injection is much sharper, shifting the relevant scales toward earlier epochs. The confinement scale for $\mu_{\rm firas}$ moves to $a_c=(1.8$–$2.3)\times10^{-7}$ with higher decay rates $\Gamma_\chi=(84$–$51)\,{\rm yr^{-1}}$, while the $y$-era remains near $a_c\!\sim\!(2.3$–$3.0)\times10^{-6}$. Compared to $\alpha=0.5$, both the $\mu$- and $y$-type distortions become more sensitive to $v_{\chi c}$, reaching up to an order of magnitude larger amplitudes for relativistic decays. These results show that steeper power-law indices favor earlier and more efficient energy injection, tightening the FIRAS and CMB upper limits on $\Sigma_\chi$ and $\Gamma_\chi$.

In Appendix~\ref{app:powerlaw} we extend our analysis to power–law decay. There, we plot the corresponding spectral–distortion amplitudes $\log_{10}\mu$ and $\log_{10}y$ across the $(a_c, v_{\chi c})$ parameter space. The results demonstrate that, overall, both slow ($\alpha=0.5$) and fast ($\alpha=6$) decay regimes reproduce similar qualitative trends, with $\mu$–type distortions dominating at early epochs and $y$–type distortions emerging later, thereby validating the exponential model as an effective approximation of more general power–law decay behavior. However, there is a dependence on $\alpha$ that modifies the duration and timing of the energy injection. In particular, large $\alpha$ values enhance the early $\mu$-distortion regime and strengthen the upper-limit constraints on $\Sigma_\chi$ for a given $\Gamma_\chi$.

\begin{figure}[t!]
\begin{center}
    \includegraphics[width=0.45\textwidth]{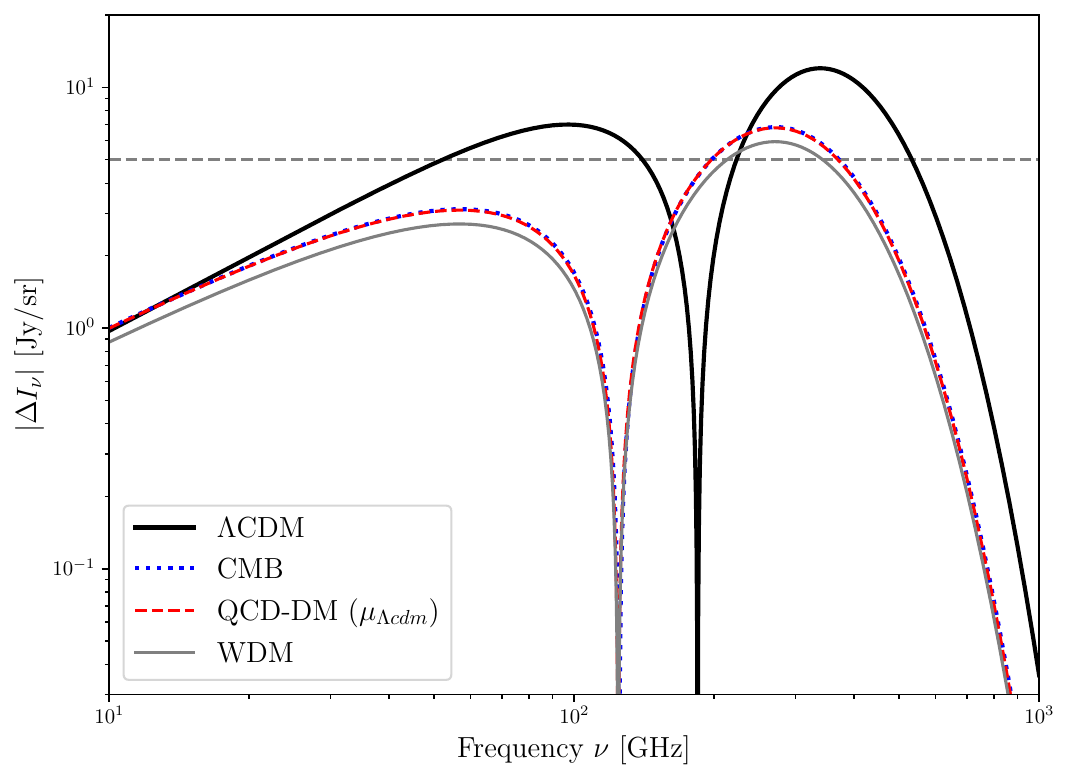}
    \caption{The intensity of CMB spectral distortion $|\Delta I_\nu|$. All cases are computed with $dT/T = 0$ to isolate the contribution of pure spectral distortions arising from QCD-DM energy injection processes. The black curve corresponds to the canonical $\Lambda$CDM prediction \{$\mu_{\Lambda \rm cdm}$, $y_{\Lambda \rm cdm}$\}, while the blue line represents a QCD-DM scenario consistent with current CMB constraints \{$\mu_{\rm firas}$, $y = 1.02 \times 10^{-7}$\}. The red curve shows a QCD-DM configuration that reproduces \{$\mu = \mu_{\Lambda \rm cdm}$, and $y \approx 0$ \}; and the gray curve corresponds to the benchmark case for a 3 keV WDM with \{$\mu = 1.43 \times 10^{-8}$ and $y\approx 0$\}. The horizontal dashed line at $|\Delta I_\nu| = 5$ Jy/sr corresponds to the expected sensitivity limit of the PIXIE experiment.}
    \label{fig:freq_vs_intensity}
\end{center}
\end{figure}

To illustrate the observational relevance of our results, Figure~\ref{fig:freq_vs_intensity} presents the absolute intensity of the CMB spectral distortion, $|\Delta I_\nu|$, for various QCD–DM benchmark scenarios, assuming $dT/T = 0$ to isolate non-thermal distortions. The black curve represents the canonical $\Lambda$CDM prediction, with distortions ${\mu_{\Lambda \rm cdm}, y_{\Lambda \rm cdm}}$ well below current detection thresholds. The blue line corresponds to a QCD–DM configuration that saturates the FIRAS $\mu$-bound while producing a suppressed $y$-type distortion ($y = 1.02 \times 10^{-7}$), compatible with Planck constraints. The red curve illustrates a case where the $\mu$-distortion matches the standard $\Lambda$CDM value, but with negligible $y$ signal, corresponding to early and brief energy injection. The gray curve shows the expected distortion for a scenario that mimics a 3 keV warm dark matter particle, yielding $\mu = 1.43 \times 10^{-8}$ and an effectively null $y$-type distortion. Notably, only the blue curve crosses the expected detection threshold of the PIXIE mission (horizontal dashed line at $|\Delta I_\nu| = 5$ Jy/sr), demonstrating that while standard and WDM-like scenarios remain undetectable, QCD–DM models with moderate decay rates and relativistic velocities can produce observable signals in the next generation of experiments.

%%% --------------------------------------  %%%
\section{Conclusions}\label{sec.conclusion} %%%
%%% --------------------------------------  %%%

%%%%% ------------------  NU - INTENSITY ------------%%%
%%%%% ------------------  NU - INTENSITY ------------%%%

The main objective of this work was to test the viability of the QCD--DM model by exploring its parameter space under the confinement-scale bounds derived from current CMB and BAO constraints. Within our fluid approximation framework, we modeled the energy injection from decaying dark-sector particles and analyzed both fast and slow decay regimes, in relativistic and non-relativistic limits. This approach provides a unified description of exponential, power-law, oscillatory, and two-step chain decays, enabling a direct comparison of their respective impacts on the $\mu$-- and $y$--type spectral distortions of the CMB.

Our results show that the numerical values obtained from exponential and power-law decays are of the same order of magnitude, confirming that the power-law, oscillatory, and two-step decay chains can be effectively modeled within the exponential framework after appropriate renormalizations. Spectral distortions, therefore, represent a robust diagnostic tool for energy-injection processes across a wide range of dark-sector dynamics, see Table~\ref{tab:decay_osc_2step}.

The impact of the decay velocity becomes relevant only in the ultra-relativistic limit ($v_{\chi c}\!\sim\!1$), while for non-relativistic scenarios ($v_{\chi c}\!\ll\!1$), the resulting spectral distortions are practically identical.  
This confirms that the main discriminant for CMB distortions lies in the decay epoch ($a_c$) and lifetime ($\Gamma_\chi$), rather than in the detailed kinematics of the decaying particles.

Assuming a decay rate of $\Gamma_\chi = 4H_c$ provides an effective benchmark for constraining the effective energy deposition parameter $\Sigma_\chi$ across the $(a_c, v_{\chi c})$ space. FIRAS bounds impose upper limits for $\mu$-type distortions at early times, which dominate over $y$-type constraints. The $\mu$ distortion sets $\log_{10}\Sigma_\chi \lesssim -2.37$ for the CMB limit $a_c = 2.66\times10^{-6}$. This emphasizes the sensitivity of SD to early energy injection.
    
The degeneracy between $\Gamma_\chi$ and $\Sigma_\chi$ implies that slower decays (small $\Gamma_\chi$) can be compensated by larger $\Sigma_\chi$ and vice versa. However, the combination of $\mu$ and $y$ constraints allows this degeneracy to be broken, especially when both observables are simultaneously matched (see Fig.~\ref{fig:gamma_sigma_ac}). The allowed region for fast decays is bounded by $\Gamma_\chi \lesssim 6.5~\text{yr}^{-1}$ and $\Sigma_\chi \lesssim 3.7 \times 10^{-4}$, beyond which spectral distortions become negligible due to efficient thermalization or observational insensitivity. 

Best-fit models simultaneously reproducing both $\mu_{\rm firas}$ and $y_{\rm firas}$ indicate that decay rates of a relativistic particle $v_{\chi c}\to1$ cluster around $\Gamma_\chi \lesssim (3.3\!-\!4.4)\times10^{-3}~\text{yr}^{-1}$, regardless of $a_c$, while $\Sigma_\chi$ decreases with increasing $a_c$ (e.g., from $1.45\times10^{-3}$ at $a_c=10^{-7}$ to $8.47\times10^{-4}$ at $a_c=2.66\times10^{-6}$). For non-relativistic decays ($v_{\chi c} \to 0$), the best solution at the CMB limit gives $\Gamma_\chi = 4\times10^{-3}~\text{yr}^{-1}$ and $\Sigma_\chi \lesssim 3.2 \times 10^{-7}$, placing the model in the slow-decay regime and highlighting that prolonged energy injection requires smaller coupling strengths to saturate FIRAS bounds.
    
In the extended analysis of power-law decays (Appendix~\ref{app:powerlaw}), both slow ($\alpha=0.5$) and fast ($\alpha=6$) decay models show similar trends to the exponential case, reinforcing its use as a general phenomenological approximation. However, steep power laws tighten the upper bounds on $\Sigma_\chi$ for a given $\Gamma_\chi$ and shift the dominant distortion epochs to earlier epochs.

Finally, our findings emphasize that upcoming missions such as {\it PIXIE} and {\it PRISM}, with sensitivities well beyond those of FIRAS, could probe $\log_{10}\Sigma_\chi\!\lesssim\!-4$ and explore regions of the $(\Gamma_\chi,\Sigma_\chi,a_c)$ parameter space that remain unconstrained by current CMB data. Such measurements would allow a direct test of the QCD--DM scenario and potentially reveal signatures of non-standard confinement dynamics in the early Universe.

\bibliographystyle{unsrtnat}
\bibliography{bd_sd}

\appendix
\section*{Acknowledgements}
JM acknowledges Investigadoras e Investigadores por México for financial support and MCTP/UNACH as the hosting institution of the IxM program. RH acknowledges financial support from UNACH.

%%%%%------------------------ ac - vxc POWER-LAW ------%%
\begin{figure*}[h!]
\centering
\begin{tabular}{cc}
\includegraphics[width=0.47\textwidth]{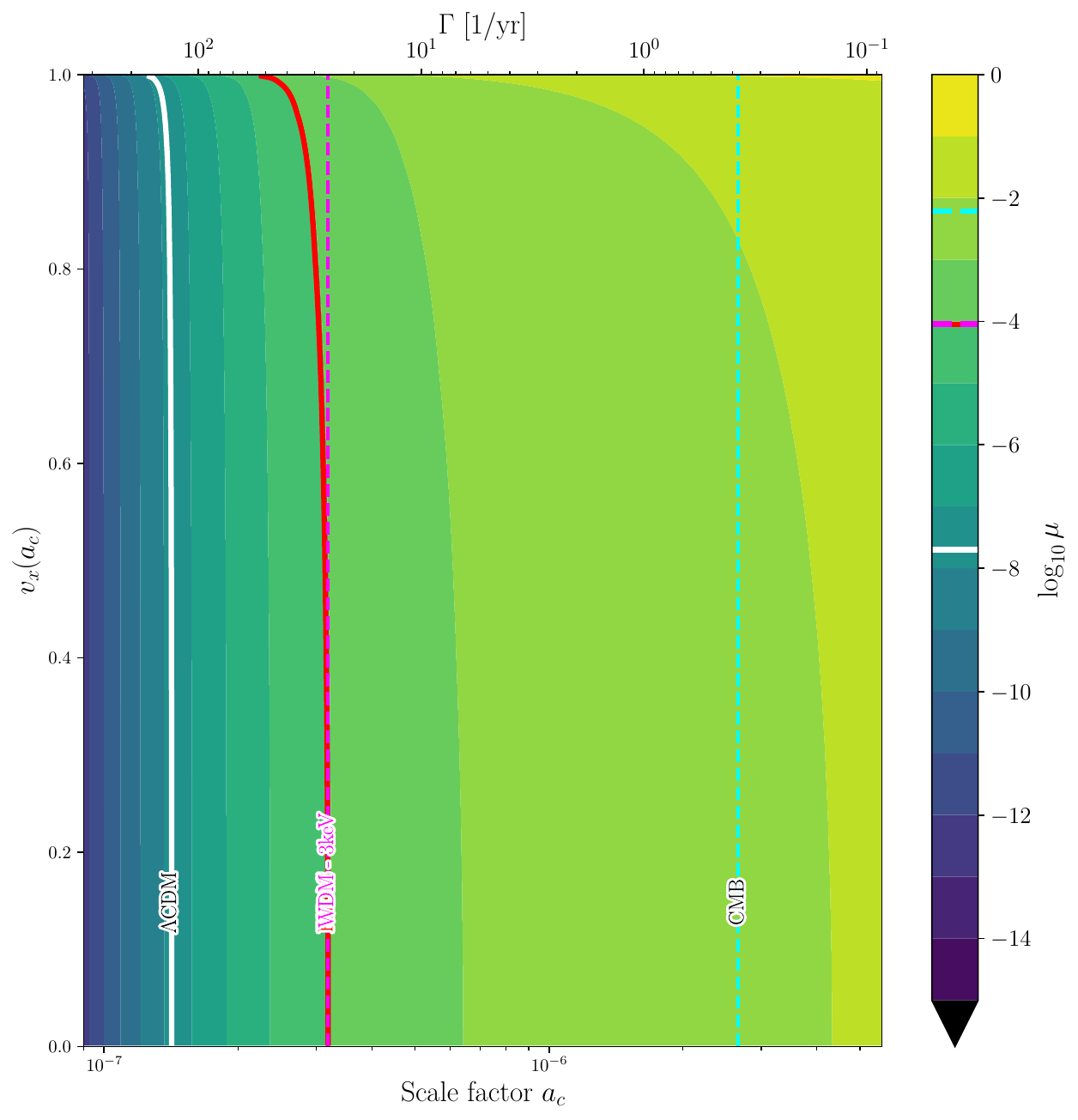} &
\includegraphics[width=0.47\textwidth]{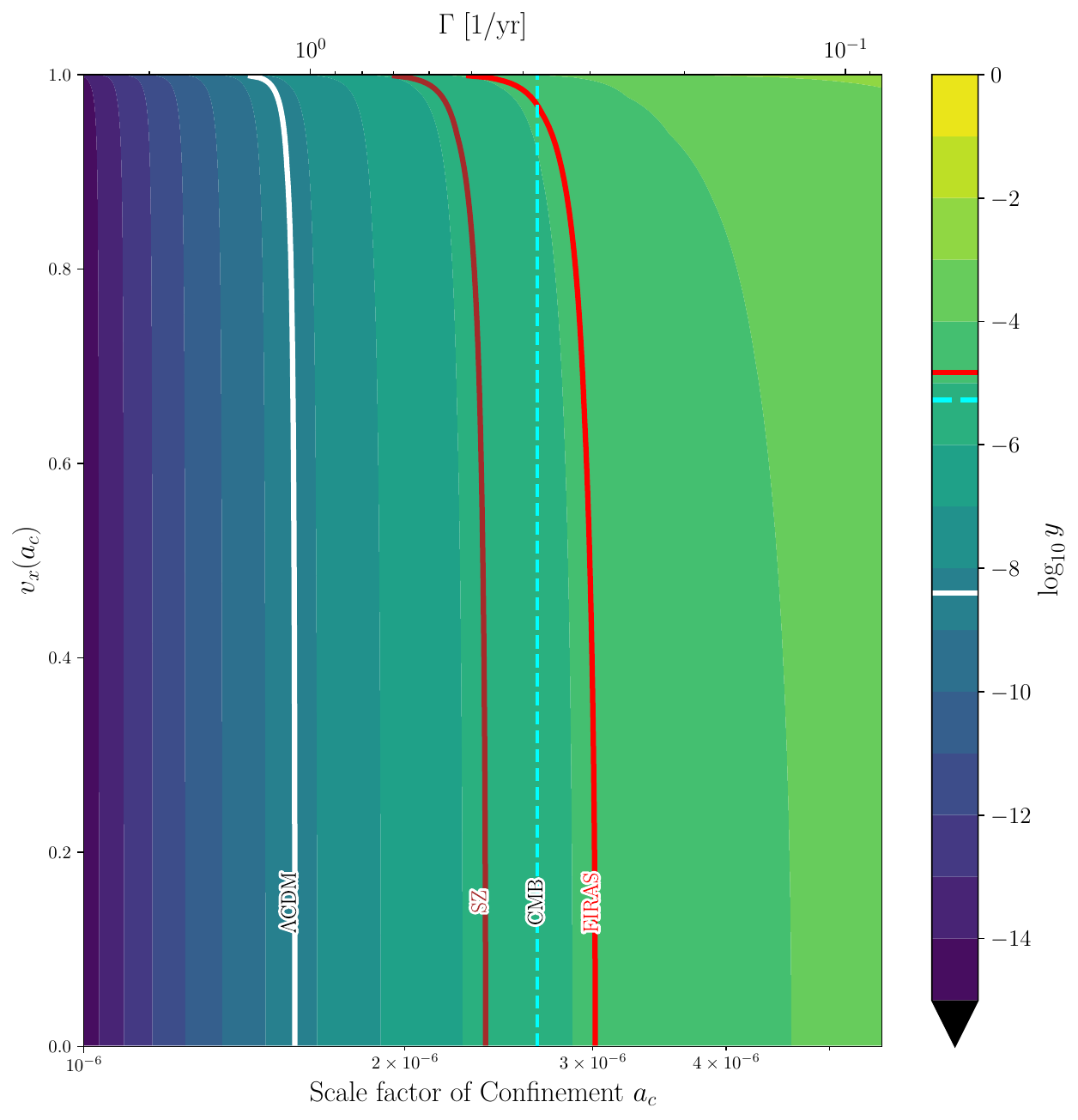} \\
(a) $\log_{10}\mu$–type SD for $\alpha=0.5$ &
(b) $\log_{10}y$–type SD for $\alpha=0.5$ \\[6pt]
\includegraphics[width=0.47\textwidth]{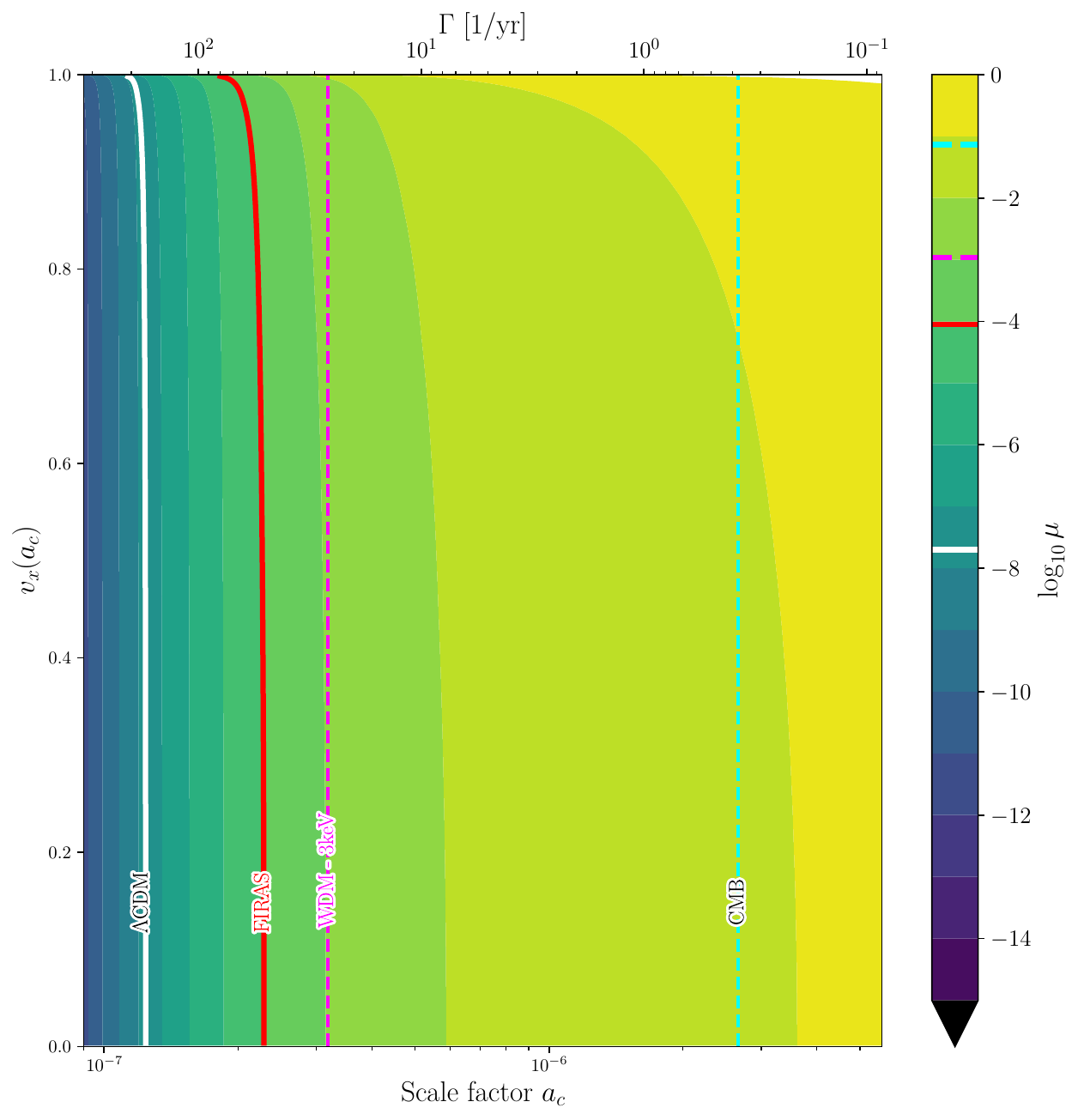} &
\includegraphics[width=0.47\textwidth]{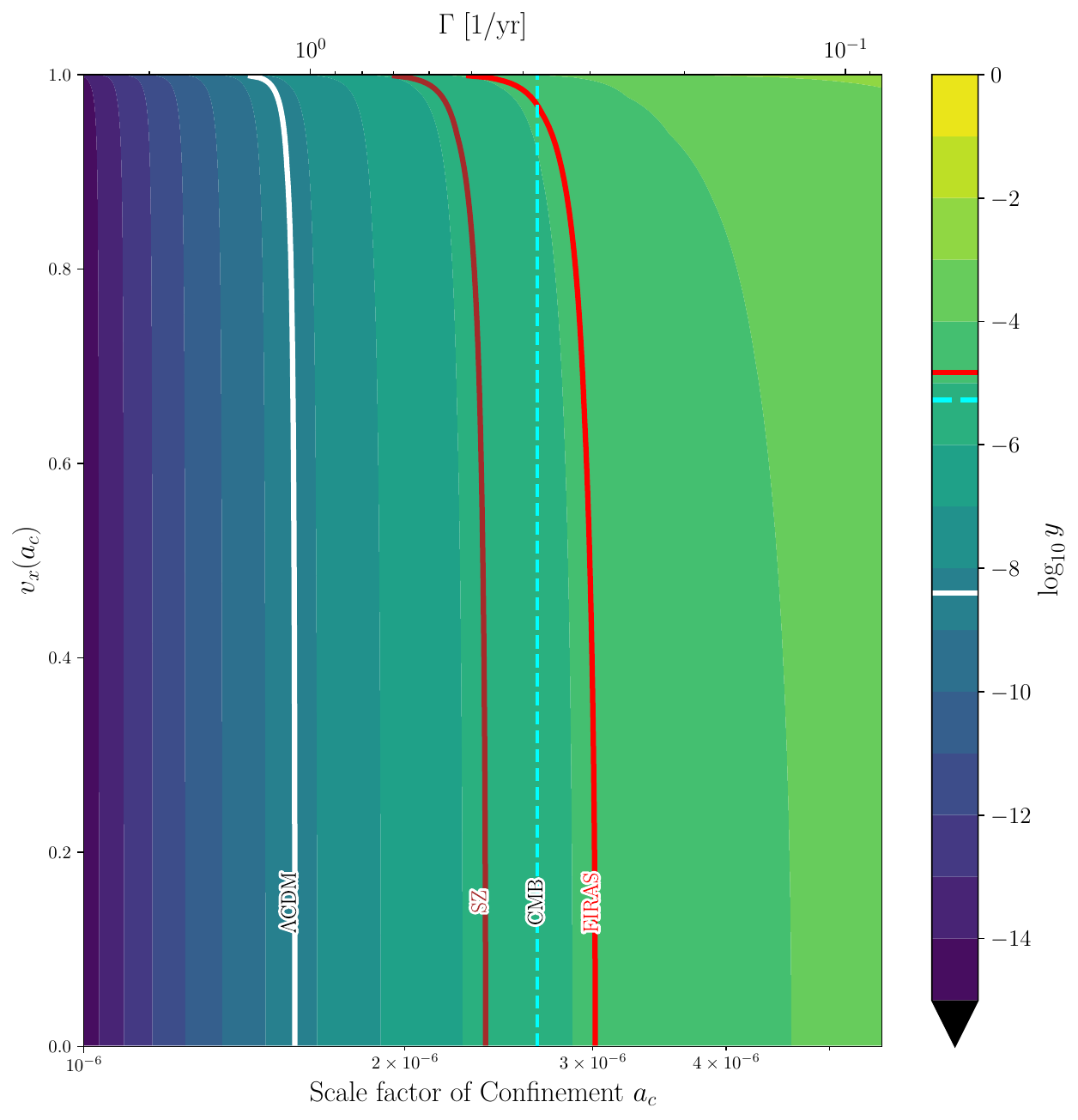} \\
(c) $\log_{10}\mu$–type SD for $\alpha=6$ &
(d) $\log_{10}y$–type SD for $\alpha=6$
\end{tabular}
\caption{
Contour maps of the spectral–distortion amplitudes in the $(a_c,v_{\chi c})$ parameter space for power–law energy injection models with $\alpha=0.5$ and $\alpha=6$.  
The color scales represent $\log_{10}(\mu/\mu_{\mathrm{FIRAS}})$ (left panels) and $\log_{10}(y/y_{\mathrm{FIRAS}})$ (right panels).  
These quantities can be directly interpreted as $\log_{10}\Sigma_\chi$ constraints normalized to FIRAS limits.  
The transition scale factor $a_c$ controls the epoch of confinement, while $v_{\chi c}$ denotes the particle velocity at decay.  
Earlier transitions ($a_c\lesssim10^{-7}$) yield stronger $\mu$–type distortions, whereas later transitions produce predominantly $y$–type signatures.  
The numerical benchmarks for each regime are provided in Table~\ref{tab:powerlaw_summary}.
}
\label{fig:powerlaw_05}
\end{figure*}
%%%%%------------------------ ac - vxc POWER-LAW ------%%

\section{Power–Law Decay Models and Spectral–Distortion Parameter Space}
\label{app:powerlaw}

In this Appendix, we present the results for the power–law decay models characterized by decay indices $\alpha=0.5$ and $\alpha=6$. For each case, we computed the logarithmic spectral–distortion amplitudes $\log_{10}\mu$ and $\log_{10}y$ over the parameter space $(a_c, v_{\chi c})$. The resulting color maps (Figures~\ref{fig:powerlaw_05}) illustrate the dependence of the $\mu$– and $y$–type distortions on the confinement scale $a_c$ and on the velocity of the decaying component $v_{\chi c}$.  

The color scales correspond to the logarithm of the normalized distortion amplitudes, $\log_{10}(\mu/\mu_{\mathrm{FIRAS}})$ and $\log_{10}(y/y_{\mathrm{FIRAS}})$. Hence, these plots can be directly interpreted as effective $\log_{10}\Sigma_\chi$ constraints: values below zero correspond to models that exceed the FIRAS bounds (requiring smaller energy fractions). In contrast, positive values indicate models that naturally fall below the current observational limit.

The parameter ranges and benchmark values associated with these maps are summarized in Table~\ref{tab:powerlaw_summary}.  
Both $\alpha=0.5$ (slow decay) and $\alpha=6$ (fast decay) exhibit similar qualitative behavior, with $\mu$–type distortions dominating the early–time regime ($a_c\lesssim10^{-7}$) and $y$–type distortions emerging at later epochs ($a_c\sim10^{-6}$–$10^{-5}$).  
The results confirm that the exponential framework can accurately approximate the power–law decay phenomenology under appropriate normalization, supporting its use as a general description of dark–sector energy injection.

\section{Dependence of $\mu$-type Spectral Distortions on the Decay Rate $\Gamma_\chi$}\label{app:decaying_rate}

\begin{figure*}[h!]
    \centering
    \includegraphics[width=0.3\textwidth]{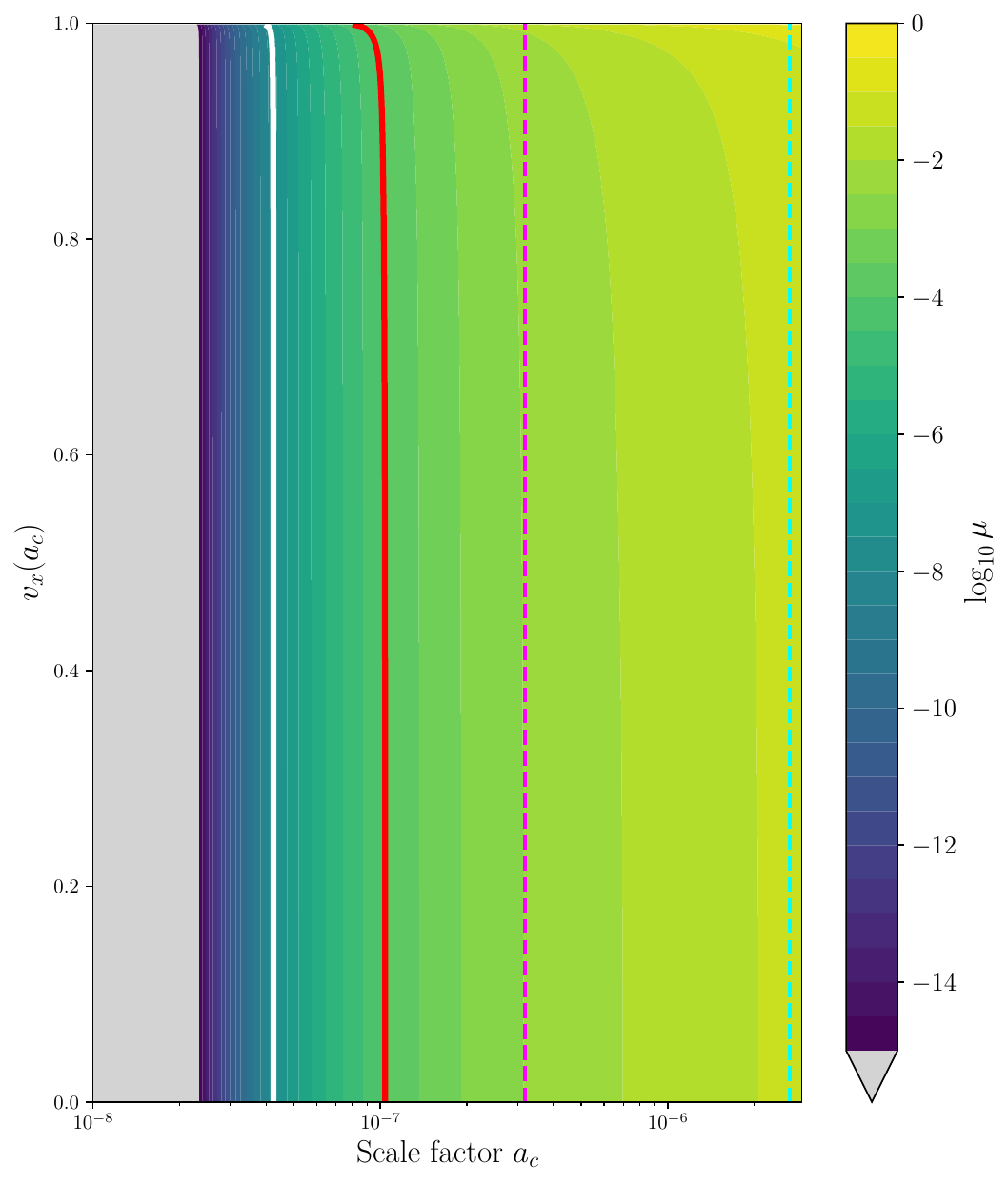}
    \includegraphics[width=0.3\textwidth]{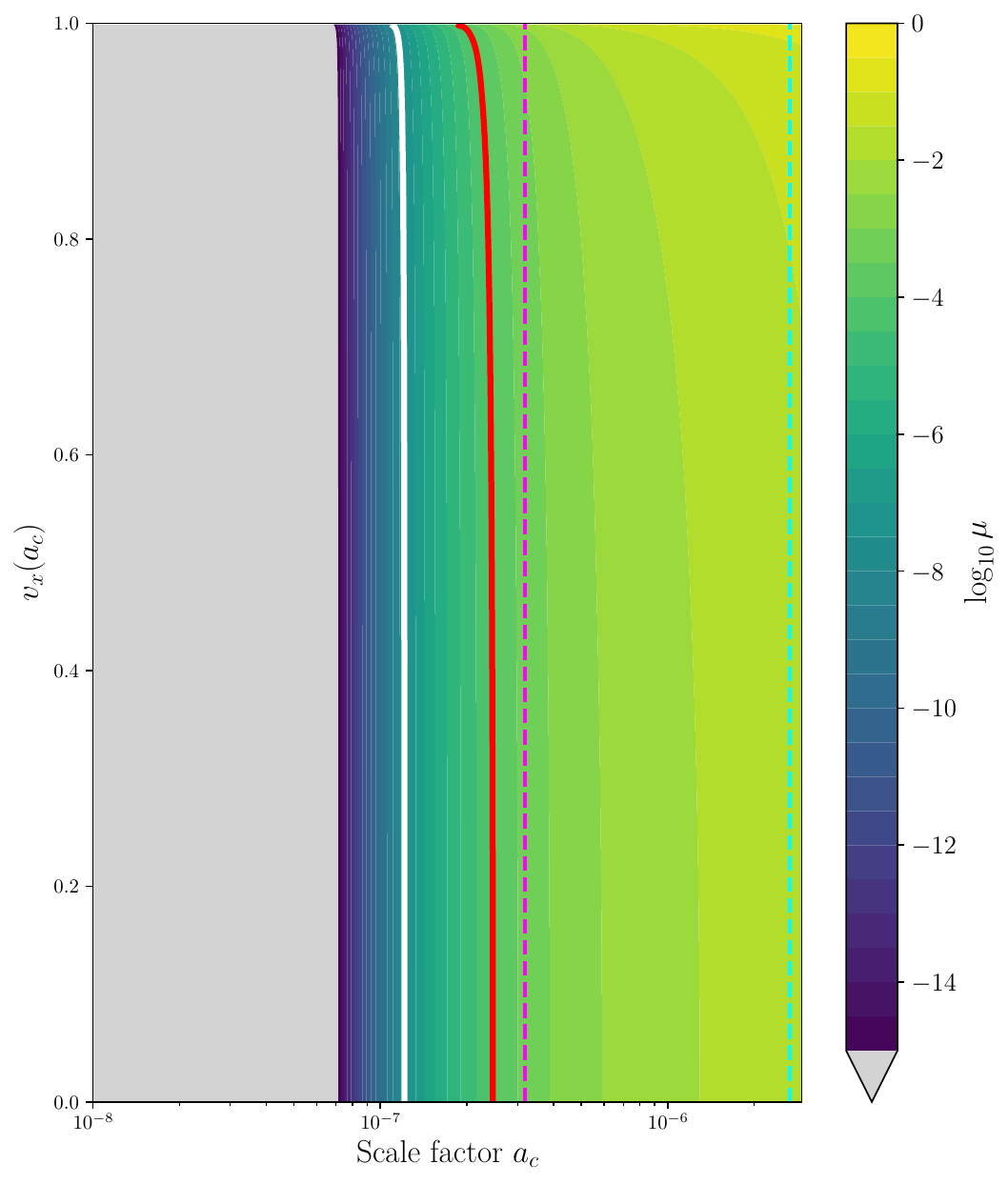}
    \includegraphics[width=0.3\textwidth]{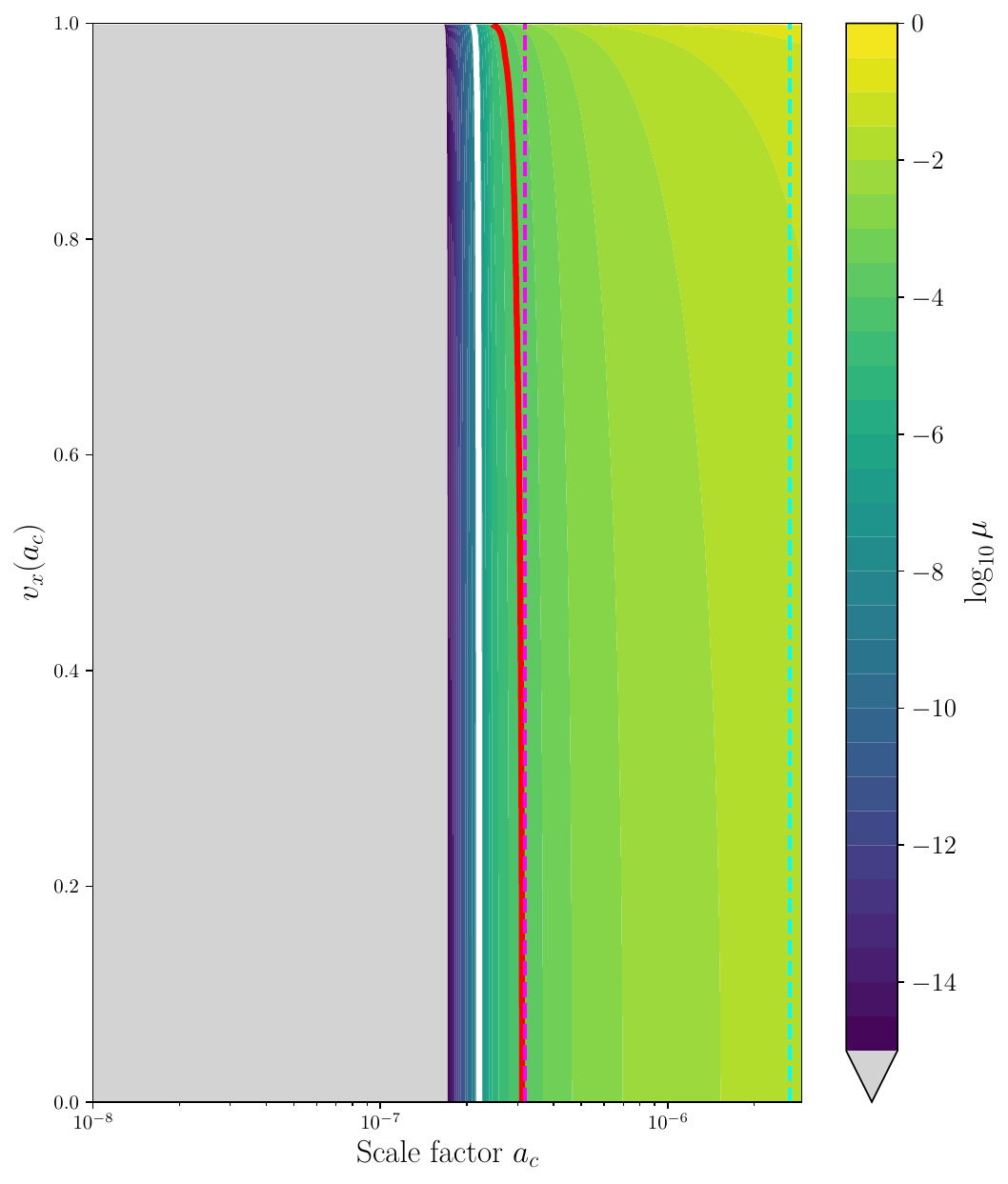}
    \caption{
    Spectral distortion amplitude of $\mu$-type shown as $\log_{10}(\mu)$ in the $(a_c, v_c)$ parameter space for a fixed decay rate $\Gamma_\chi = X \times 4H_c$, where $X = 0.1$ (left), $1$ (middle), and $10$ (right). 
    The gray regions correspond to values below the observational sensitivity threshold $\mu < 10^{-15}$.
    The red contour denotes the upper limit from FIRAS, while the white contour indicates the expected $\mu$ value from the standard $\Lambda$CDM model.
    The vertical cyan line shows the upper bound on $a_c$ from Planck CMB observations for the stable dark sector scenario.
    All other model parameters are held fixed, including $\Sigma_\chi = 1$. These panels illustrate the linear dependence of the spectral distortion amplitude on $\Gamma_\chi$, with larger decay rates yielding more visible distortions. Notably, the extent of the region producing detectable distortions shifts with $\Gamma_\chi$.
    }
    \label{fig:mu_decay_variation}
\end{figure*}

Spectral distortion amplitude of $\mu$-type shown as $\log_{10}(\mu)$ in the $(a_c, v_c)$ parameter space for a fixed decay rate $\Gamma_\chi = X \times 4H(a_c)$, where $X = 0.1$ (top), $1$ (middle), and $10$ (bottom). 
The gray regions correspond to values below the observational sensitivity threshold $\mu < 10^{-15}$.
The red contour denotes the upper limit from FIRAS, while the white contour indicates the expected $\mu$ value from the standard $\Lambda$CDM model.
The vertical cyan line shows the upper bound on $a_c$ from Planck CMB observations for the stable dark sector scenario. All other model parameters are held fixed, including $\Sigma_\chi = 1$. These panels illustrate the linear dependence of the spectral distortion amplitude on $\Gamma_\chi$, with larger decay rates yielding more visible distortions. Notably, the extent of the region producing detectable distortions shifts with $\Gamma_\chi$.

In this appendix, we show how the choice of the decay rate $\Gamma_\chi$ affects the predicted $\mu$-type SDs in the $(a_c, v_c)$ parameter space. This analysis provides a complementary view to the main results by explicitly showing the sensitivity of $\mu$-distortions to the decay time of the unstable dark particle $\chi$.

Figure~\ref{fig:mu_decay_variation} presents three panels corresponding to fixed decay rates $\Gamma_\chi = 0.1 \times 4H(a_c)$, $1 \times 4H(a_c)$ (equivalent to Fig.~\ref{fig:log10_sigma_mu}), and $10 \times 4H(a_c)$, respectively. The quantity $\log_{10}(\mu)$ is plotted across the $(a_c, v_c)$ parameter space, assuming all other model parameters remain constant, including the effective energy injection fraction $\Sigma_\chi = 1$.

Gray-shaded regions represent values of $\mu$ below the detectability threshold $\mu < 10^{-15}$, while the red contour indicates the FIRAS upper limit on $\mu$-distortions. The vertical cyan line marks the upper bound on the confinement scale factor $a_c$ from Planck CMB data under the assumption of a stable dark sector. The white contour shows the $\mu$-distortion amplitude expected in the standard $\Lambda$CDM model.

These plots demonstrate that the spectral distortion amplitude scales approximately linearly with the decay rate $\Gamma_\chi$. In particular, the transition between visible and undetectable distortions occurs at different regions in parameter space depending on the value of $\Gamma_\chi$. Faster decays ($\Gamma_\chi \gg H$) inject energy promptly and more efficiently into the photon bath, resulting in more pronounced distortions. Conversely, for slow decays ($\Gamma_\chi \ll H$), the energy injection occurs later, when the thermalization efficiency is reduced, suppressing the $\mu$ signal.

\end{document}